\documentclass[sigconf]{acmart}

\usepackage[textsize=tiny]{todonotes}
\usepackage{color}
\usepackage{graphicx}
\usepackage{threeparttable}
\usepackage{amsmath,amssymb,amsfonts,amsthm}
\usepackage{subfigure}
\usepackage{multirow} 
\usepackage{wrapfig}
\usepackage{amsmath,bm}
\usepackage{xpatch}
\usepackage{enumitem}
\usepackage{caption}
\usepackage{algorithm}  
\usepackage{algorithmic}
\usepackage{hyperref}
\hypersetup{hidelinks,
	colorlinks=true,
	linkcolor=black,
	allcolors=black,
	pdfstartview=Fit,
	breaklinks=true}

\newcommand{\model}{\text{PANE-GNN}}
\newcommand{\like}{\text{interest}}
\newcommand{\dislike}{\text{disinterest}}

\AtBeginDocument{%
  \providecommand\BibTeX{{%
    \normalfont B\kern-0.5em{\scshape i\kern-0.25em b}\kern-0.8em\TeX}}}

\setcopyright{acmcopyright}
\copyrightyear{2023}
\acmYear{2023}
\acmDOI{XXXXXXX.XXXXXXX}

\acmConference[CIKM '23]{The 32nd ACM International Conference on Information and Knowledge Management}{October 21-25, 2023}{Birmingham, UK}

\acmPrice{15.00}
\acmISBN{978-1-4503-XXXX-X/18/06}

\begin{document}

% \title{PANE-GNN: Unifying Positive and Negative Edges \\
% in Graph Neural Networks for Recommendation}

\title{(Technical Report) \\
PANE-GNN: Unifying Positive and Negative Edges \\
in Graph Neural Networks for Recommendation}

\author{Ziyang Liu$^{\dagger}$, Chaokun Wang$^{\dagger}$, Jingcao Xu$^{\dagger}$, Cheng Wu$^{\dagger}$, Kai Zheng$^{\star}$, Yang Song$^{\star}$}
\author{Na Mou$^{\star}$, Kun Gai$^{\diamondsuit}$}
\email{{liu-zy21,xjc20,wuc22}@mails.tsinghua.edu.cn,chaokun@tsinghua.edu.cn}
\email{{zhengkai,mouna}@kuaishou.com,	ys@sonyis.me,gai.kun@qq.com}
% \authornotemark[1]
\affiliation{%
  \institution{School of Software, Tsinghua University$^{\dagger}$  Kuaishou$^{\star}$
  Unaffilited$^{\diamondsuit}$}
  \city{Beijing}
  \country{China}
}

% \author{}
% % \authornote{Both authors contributed equally to this research.}
% \email{zhengkai@kuaishou.com}
% % \authornotemark[1]
% \affiliation{%
%   \institution{Kuaishou}
%   \city{Beijing}
%   \country{China}
% }

% \author{\IEEEauthorblockN{Ziyang Liu,
% Chaokun Wang,
% Yunkai Lou,
% Hao Feng}
% \IEEEauthorblockA{School of Software, Tsinghua University, Beijing 100084, China\\
% Email: \{liu-zy21, louyk18, fh20\}@mails.tsinghua.edu.cn, chaokun@tsinghua.edu.cn}}

\begin{abstract}
    Recommender systems play a crucial role in addressing the issue of information overload by delivering personalized recommendations to users. In recent years, there has been a growing interest in leveraging graph neural networks (GNNs) for recommender systems, capitalizing on advancements in graph representation learning. These GNN-based models primarily focus on analyzing users' positive feedback while overlooking the valuable insights provided by their negative feedback.
    In this paper, we propose $\model$, an innovative recommendation model that unifies \textbf{P}ositive \textbf{A}nd \textbf{N}egative \textbf{E}dges in \textbf{G}raph \textbf{N}eural \textbf{N}etworks for recommendation. By incorporating user preferences and dispreferences, our approach enhances the capability of recommender systems to offer personalized suggestions.
    $\model$ first partitions the raw rating graph into two distinct bipartite graphs based on positive and negative feedback. Subsequently, we employ two separate embeddings, the $\like$ embedding and the $\dislike$ embedding, to capture users' likes and dislikes, respectively. To facilitate effective information propagation, we design distinct message-passing mechanisms for positive and negative feedback.
    Furthermore, we introduce a distortion to the negative graph, which exclusively consists of negative feedback edges, for contrastive training. This distortion plays a crucial role in effectively denoising the negative feedback.
    The experimental results provide compelling evidence that $\model$ surpasses the existing state-of-the-art benchmark methods across four real-world datasets. These datasets include three commonly used recommender system datasets and one open-source short video recommendation dataset.
\end{abstract}




\maketitle

\pagestyle{plain} %显示页码

\section{Introduction}
\label{Introduction}
% recommender system, CF, & GNNs
Recommender systems have garnered significant attention as a prominent research field, offering solutions in information filtering by predicting users' item ratings or preferences~\cite{cikm-rs-1,cikm-rs-2,cikm-rs-3,cikm-rs-4,BPRMF,NeuMF}. The versatility of recommender systems is evident through their widespread adoption across diverse domains, establishing their importance in recent years. Notably, recommender systems have found practical applications in domains such as movies~\cite{movie-1,movie-2}, news~\cite{news-1,news-2}, e-commerce items~\cite{item-1,cikm-rs-3}, and short videos~\cite{short-video-1,short-video-2}. Given the multitude of domains and the increasing reliance on these systems, the development of recommender systems has emerged as a critical concern within the field of computer science~\cite{cf-survey,2023_rs_1,2023_rs_2}.

\begin{figure}[b]
\setlength{\abovecaptionskip}{-1mm} 
\begin{center}
\includegraphics[width=0.41\textwidth]{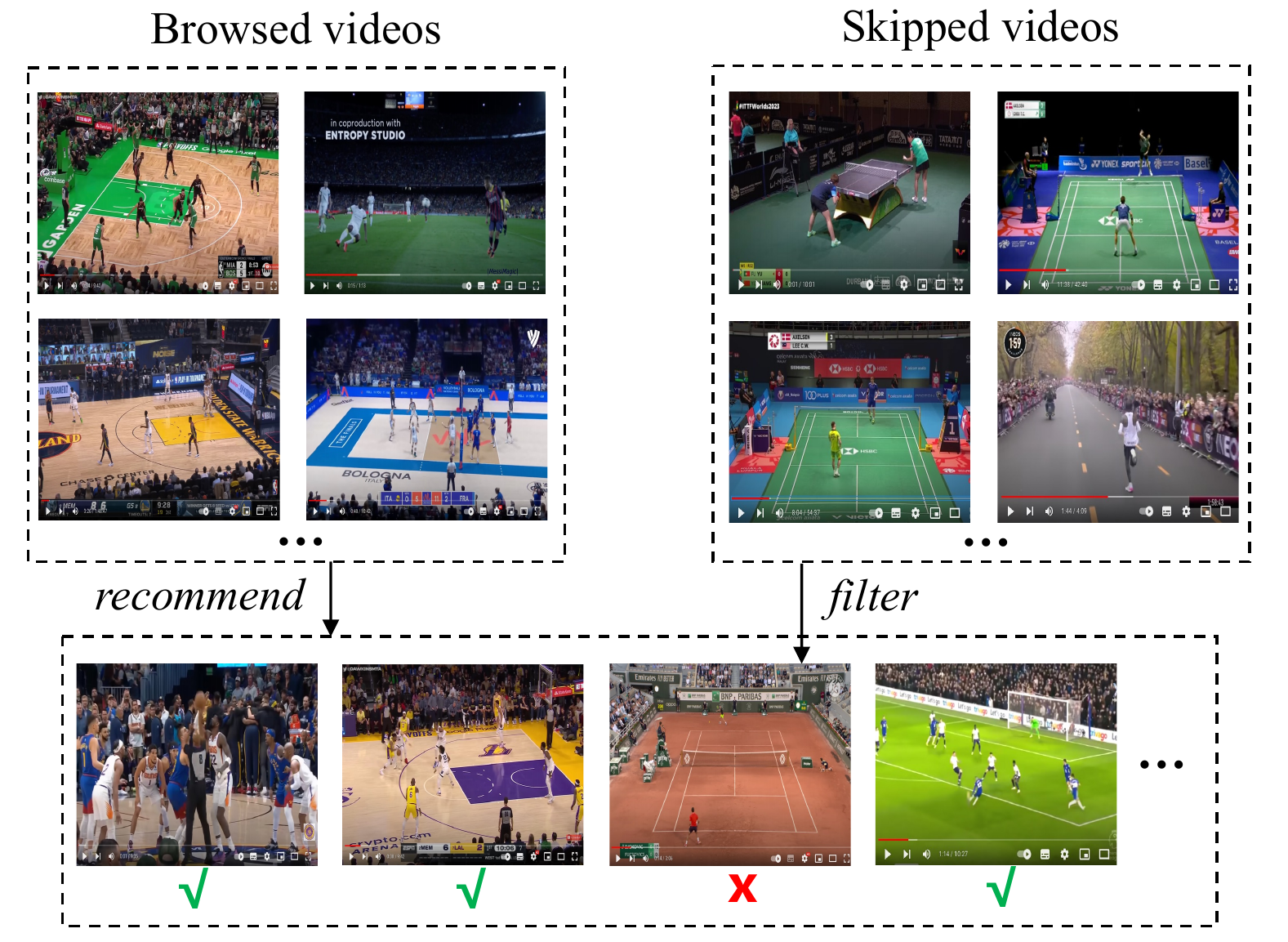}
\end{center}
\caption{An example of video recommendation from YouTube. The integration of positive and negative feedback plays a pivotal role in achieving accurate recommendation outcomes. In this example, the user prefers team sports while showing no interest in single-player sports.}
\label{intro-model}
\end{figure}

The fundamental structure of recommender systems' user-item interaction graph can be represented as a signed bipartite graph, encompassing both positive and negative feedback from users. Positive feedback indicates user interest, while negative feedback denotes disinterest or dissatisfaction. Major platforms such as YouTube and Amazon offer mechanisms for users to express their preferences or assign ratings to items, reflecting these feedback categories.
The incorporation of negative feedback assumes significance in cases where positive feedback signals are absent, serving as a critical means to prevent irrelevant or unnecessary recommendations to users. Figure~\ref{intro-model} visually demonstrates the integration of positive and negative feedback, highlighting its potential for accurate recommendation outcomes.

\textbf{Observation 1}:
Existing graph neural network (GNN) paradigms for recommendations fail to effectively incorporate negative feedback information, particularly in the context of message passing~\cite{gnn4rs-1,gnn4rs-2,gnn4rs-3,lightgcn,siren}. This limitation hampers the comprehensive utilization of valuable user feedback in recommender systems.

\begin{figure}[t]
\centering
\setlength{\abovecaptionskip}{-0mm}
\setlength{\belowcaptionskip}{-0mm}
\hspace{0.0in}
\subfigure[\scriptsize{\textcolor{black}{Precision results.}}]{
\includegraphics[width=0.14\textwidth]{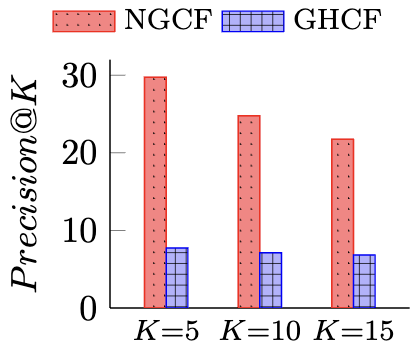}
\hspace{0.0in}}
\subfigure[\scriptsize{\textcolor{black}{Recall results.}}]{
\includegraphics[width=0.14\textwidth]{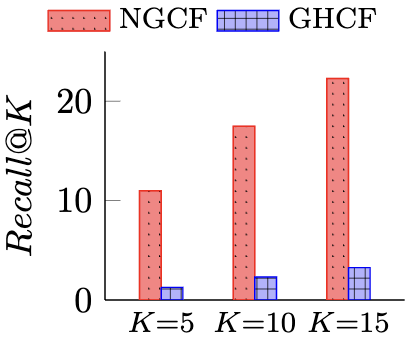}
\hspace{0.0in}}
\subfigure[\scriptsize{\textcolor{black}{nDCG results.}}]{
\includegraphics[width=0.14\textwidth]{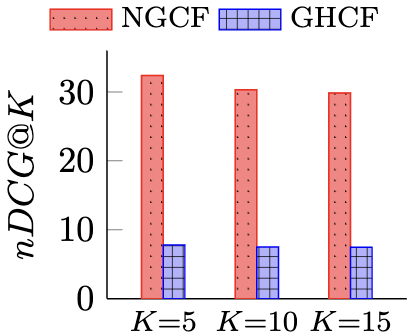}
\hspace{0.0in}}
\caption{\textcolor{black}{Comparison of single-relational (NGCF) and multi-relational (GHCF) recommendation models on the ML-1M dataset.}}
\label{comp}
\vspace*{-8mm}
\end{figure}

\textbf{Observation 2}:
While incorporating negative feedback through a multiple-behavioral graph seems like a natural approach, our experiments with the state-of-the-art method GHCF~\cite{GHCF} (shown in Figure~\ref{comp}) reveal a decrease in performance compared to NGCF that does not utilize negative feedback. It suggests that directly incorporating negative feedback may not always yield benefits.

\textbf{Challenges}.
The aforementioned observations underscore the challenge of developing effective algorithms that can effectively incorporate negative feedback into recommender systems. The underutilization of negative feedback in current approaches motivates us to explore the usage of negative feedback through GNNs in order to enhance the quality of recommendations.
However, learning high-order structural information from a signed bipartite graph  faces difficulties due to the limitations of the \textit{network homophily assumption} and the \textit{balance theory assumption}.
The network homophily assumption posits that similar nodes are more likely to connect to each other than dissimilar nodes. Many GNN models~\cite{NGCF,LR-GCCF,lightgcn} adopt a message-passing mechanism that aggregates information from local neighbors to update the embedding of the anchor node based on this assumption. However, homophily is not applicable in signed graphs where dissimilar nodes are connected by negative edges.
The balance theory assumption implies that ``the friend of my friend is my friend'', ``the enemy of my friend is my enemy'', and ``the enemy of my enemy is my friend''.
Existing methods for signed unipartite graphs~\cite{SGCN,signed_gnn_1,signed_gnn_2} leverage this assumption to aggregate and propagate information across layers.
However, the balance theory assumption does not match with the signed bipartite graph in recommender systems~\cite{balance_theory_1,balance_theory_2,balance_theory_3}. In real-world recommendation scenarios, users typically possess diverse interests rather than unique interests.
Consequently, the fundamental idea of ``the enemy of my enemy is my friend'' (i.e., ``two items disliked by the same user are similar'') in the balance theory assumption does not accurately capture the complexity of real-world situations.
These limitations necessitate the development of novel approaches to effectively leverage negative feedback in recommender systems, accounting for the unique characteristics of signed bipartite graphs and the diverse interests of users in real-world settings.

\textbf{Our idea}.
The key idea revolves around utilizing high-order structural information from both the positive graph (i.e., user-item interaction graph containing only positive feedback edges) and the negative graph (i.e., user-item interaction graph containing only negative feedback edges) simultaneously.
To enhance recommendations by incorporating negative feedback, this paper presents a novel recommendation model called $\model$ (unifying \textbf{P}ositive \textbf{A}nd \textbf{N}egative \textbf{E}dges in \textbf{G}raph \textbf{N}eural \textbf{N}etworks for recommendation). In this model, each user or item is assigned two embeddings, i.e., $\like$ embedding and $\dislike$ embedding, to capture the user's interests and disinterests, respectively.
Taking into account the network homophily assumption, we devise two message-passing mechanisms for the positive graph and the negative graph. On the positive graph, $\like$ embeddings are propagated and updated, capturing the user's interests. On the other hand, on the negative graph, disinterest embeddings are propagated and updated, capturing the user's disinterests or items they explicitly dislike.
Furthermore, to generate robust embeddings that remain invariant to graph perturbations, we utilize graph contrastive learning on the negative graph and its perturbed version. This approach enhances the model's ability to capture relevant patterns in the presence of graph noise.

The main three contributions of this work are as follows:
\begin{itemize}[leftmargin=*]
    \item We propose a novel GNN-based recommendation model called $\model$. The model performs message passing on both the positive graph and the negative graph to effectively incorporate positive and negative feedback (Section~\ref{Message passing on the positive graph and negative graph}).
    \item We design contrastive learning on the negative graph (Section~\ref{Contrastive learning on the positive graph}), a new ranking method with a disinterest-score filter (Section~\ref{Ranking with a disinterest-score filter}), and a dual feedback-aware Bayesian personalized ranking loss (Section~\ref{Optimization}), all of which improve recommendation accuracy through the integration of positive and negative feedback signals.
    \item The proposed $\model$ is extensively evaluated on four real-world datasets (Section~\ref{Experiment}). The experimental results demonstrate that $\model$ outperforms state-of-the-art GNN-based recommendation methods.
\end{itemize}

\section{Related Work}
\label{Related Work}
We provide a review of existing work about 1) recommender systems based on GNNs, and 2) graph neural networks on signed graphs.

\vspace*{-3mm}
\subsection{Recommender Systems based on GNNs}
\label{Recommender systems based on Graph Neural Networks}
% NGCF, LR-GCCF, LightGCN, SiReN
Recently, GNNs have become the new state-of-the-art approach in many recommendation problems~\cite{tutorial_gnn_rs,survey_gnn_rs}.
The main advantage of using GNNs for recommender systems is that it can capture higher-order structural information in the observed data.
Based on the message-passing architecture of GNNs, NGCF~\cite{NGCF} adopts the Hadamard product between user embedding and item embedding to promote passing more messages from similar items to users.
Considering that nonlinear activation contributes little to the recommendation performance, LR-GCCF~\cite{LR-GCCF} removes non-linearities from the original graph convolutional network (GCN) model~\cite{gcn} and adds a residual network structure on it to alleviate the over-smoothing problem in the graph convolution aggregation.
Likewise, LightGCN~\cite{lightgcn} removes both feature transformation and nonlinear activation and only retains neighborhood aggregation for collaborative filtering. The simplified model has higher computational efficiency and is much easier to implement and train.

Our proposed method differs from the above methods in that we consider the negative feedback information in the observed data and devise a novel message-passing process that takes into account both positive and negative feedback. 

\vspace*{-3mm}
\subsection{Graph Neural Networks on Signed Graphs}
\label{Graph Neural Networks on Signed Graphs}
% SGCN,SNEA,SigGAN,SiReN
Most of the previous work focus on building GNNs for unsigned graphs where there are only positive edges.
Currently, signed graphs where each edge has a positive or negative sign, have become increasingly ubiquitous in the real world. For example, the users in a social network may hold common or opposite political views.
Since the network homophily assumption is the theoretical basis of the message-passing mechanism in GNNs, those unsigned GNNs cannot be applied to signed graphs directly.
As a pioneering work of signed GNNs, SGCN~\cite{SGCN} assigns balanced embedding and unbalanced embedding for each node and propagates the two embeddings in the signed graph based on balance theory.
Furtherly, SNEA~\cite{SNEA} optimizes the message-passing process in SGCN by assigning different importance coefficients to each node pair connected with different edges.
Inspired by adversarial learning, ASiNE~\cite{ASiNE} plays a minimax game in the positive graph and negative graph by leveraging a generator and a discriminator for positive edges and negative edges in a signed graph, respectively. 
SiReN~\cite{siren} generates positive embeddings and negative embeddings for each node in a signed graph via a GNN model and a multilayer perceptron (MLP) model, respectively. Then SiReN adopts an attention layer to integrate the two embeddings into the final embeddings.

Unlike the existing methods based on the balance theory assumption, which may not be directly applicable to the signed bipartite graph in recommender systems, the proposed method in this work takes a different approach.
It splits the raw rating graph into two distinct graphs and emphasizes the propagation of information within each graph based on the type of edges.

\section{Method}
\label{Method}
In this section, we introduce the notations used in the paper, present the architecture of $\model$, and describe its optimization objective.

\subsection{Notations}
\label{Notations}
In the given raw rating graph $\mathcal{G}=(\mathcal{U},\mathcal{I},\mathcal{E})$, where $\mathcal{U}$ represents the set of users, $\mathcal{I}$ represents the set of items, and $\mathcal{E}$ represents the set of edges, we split the graph into two edge-disjoint graphs: the positive graph $\mathcal{G}{p}=(\mathcal{U},\mathcal{I},\mathcal{E}{p})$ and the negative graph $\mathcal{G}{n}=(\mathcal{U},\mathcal{I},\mathcal{E}{n})$. Here, $\mathcal{E}{p}$ represents the edges corresponding to positive ratings, and $\mathcal{E}{n}$ represents the edges corresponding to negative ratings. The union of $\mathcal{E}{p}$ and $\mathcal{E}{n}$ gives the set of all edges $\mathcal{E}$.
In the positive graph $\mathcal{G}{p}$, we aim to learn the $\like$ embeddings for users and items, denoted as $\mathbf{z}_{u}$ and $\mathbf{z}_{i}$, respectively. These embeddings capture the relationship between liking and being liked.
In contrast, in the negative graph $\mathcal{G}{n}$, we focus on learning the $\dislike$ embeddings for users and items, represented as $\mathbf{v}_{u}$ and $\mathbf{v}_{i}$, respectively. These embeddings capture the relationship between disliking and being disliked.
For a comprehensive overview of the notations used in this paper, please refer to Table~\ref{Frequently used notations in this paper}.

\begin{table}[t]
\setlength{\abovecaptionskip}{-0mm} 
\setlength{\belowcaptionskip}{-0mm}
\setlength\tabcolsep{2.0pt}
\small
\caption{Frequently used notations in this paper.}
\label{Frequently used notations in this paper}
\begin{center}
\begin{threeparttable}
\begin{tabular}{cc}
\toprule[1pt]
\bf Notation &\bf Description\\
\midrule[0.5pt]
$\mathcal{U}$ &Set of users. \\
$\mathcal{I}$ &Set of items. \\
$\mathcal{E}_{p}$ &Set of positive edges. \\
$\mathcal{E}_{n}$ &Set of negative edges. \\
$\mathcal{E}=\mathcal{E}_{p}\cup \mathcal{E}_{n}$ &Set of all edges. \\
$\mathcal{G}{=}(\mathcal{U}, \mathcal{I}, \mathcal{E})$ &Raw rating graph. \\
$\mathcal{G}_{p}{=}(\mathcal{U}, \mathcal{I}, \mathcal{E}_{p})$ &Positive graph. \\
$\mathcal{G}_{n}{=}(\mathcal{U}, \mathcal{I}, \mathcal{E}_{n})$ &Negative graph. \\
$\mathcal{G}_{d}{=}(\mathcal{U}, \mathcal{I}, \mathcal{E}_{d})$ &Distorted graph from $\mathcal{G}_{n}$. \\
$N=|\mathcal{U}{\cup}\mathcal{I}|$ &Number of all nodes in $\mathcal{G}$. \\
$\mathbf{A}_{p},\mathbf{A}_{n},\mathbf{A}_{d}{\in} \mathbb{R}^{N{\times} N}$ &Adjacency matrices of $\mathcal{G}_{p}$, $\mathcal{G}_{n}$, \& $\mathcal{G}_{d}$. \\
$\mathcal{N}_{p}(u),\mathcal{N}_{n}(u),\mathcal{N}_{d}(u)$ &Neighbor sets of user $u$ in $\mathcal{G}_{p}$, $\mathcal{G}_{n}$, \& $\mathcal{G}_{d}$. \\
$\mathcal{N}_{p}(i),\mathcal{N}_{n}(i),\mathcal{N}_{d}(i)$ &Neighbor sets of item $i$ in $\mathcal{G}_{p}$, $\mathcal{G}_{n}$, \& $\mathcal{G}_{d}$. \\
$\mathbf{Z}{\in} \mathbb{R}^{N\times H}$ &Interest embedding matrix. \\
$\mathbf{V}{\in} \mathbb{R}^{N\times H}$ &Disinterest embedding matrix. \\
$\mathbf{z}_{u},\mathbf{z}_{i}{\in} \mathbb{R}^{H}$ &Interest embeddings on $\mathcal{G}_{p}$. \\
$\mathbf{v}_{u},\mathbf{v}_{i}{\in} \mathbb{R}^{H}$ &Disinterest embeddings on $\mathcal{G}_{n}$. \\
$\tilde{\mathbf{v}}_{u},\tilde{\mathbf{v}}_{i}{\in} \mathbb{R}^{H}$ &Disinterest embeddings on $\mathcal{G}_{d}$. \\
\midrule[0.5pt]
$H$ &Embedding size. \\
$K$ &Layer number of graph neural networks. \\
$p$ &Probability of edge removing. \\
$b$ &Feedback-aware coefficient. \\
$\delta$ &Filtering threshold. \\
$\lambda_{1}$ &Contrastive learning coefficient. \\
$\lambda_{2}$ &L2 regularization coefficient. \\
$\tau$ &Temperature coefficient. \\
\bottomrule[1pt]
\end{tabular}
\end{threeparttable}
\end{center}
\vspace*{-3mm}
\end{table}

\begin{figure*}[t]
\setlength{\abovecaptionskip}{-0mm} 
\setlength{\belowcaptionskip}{-0mm}
\begin{center}
\includegraphics[width=0.90\textwidth]{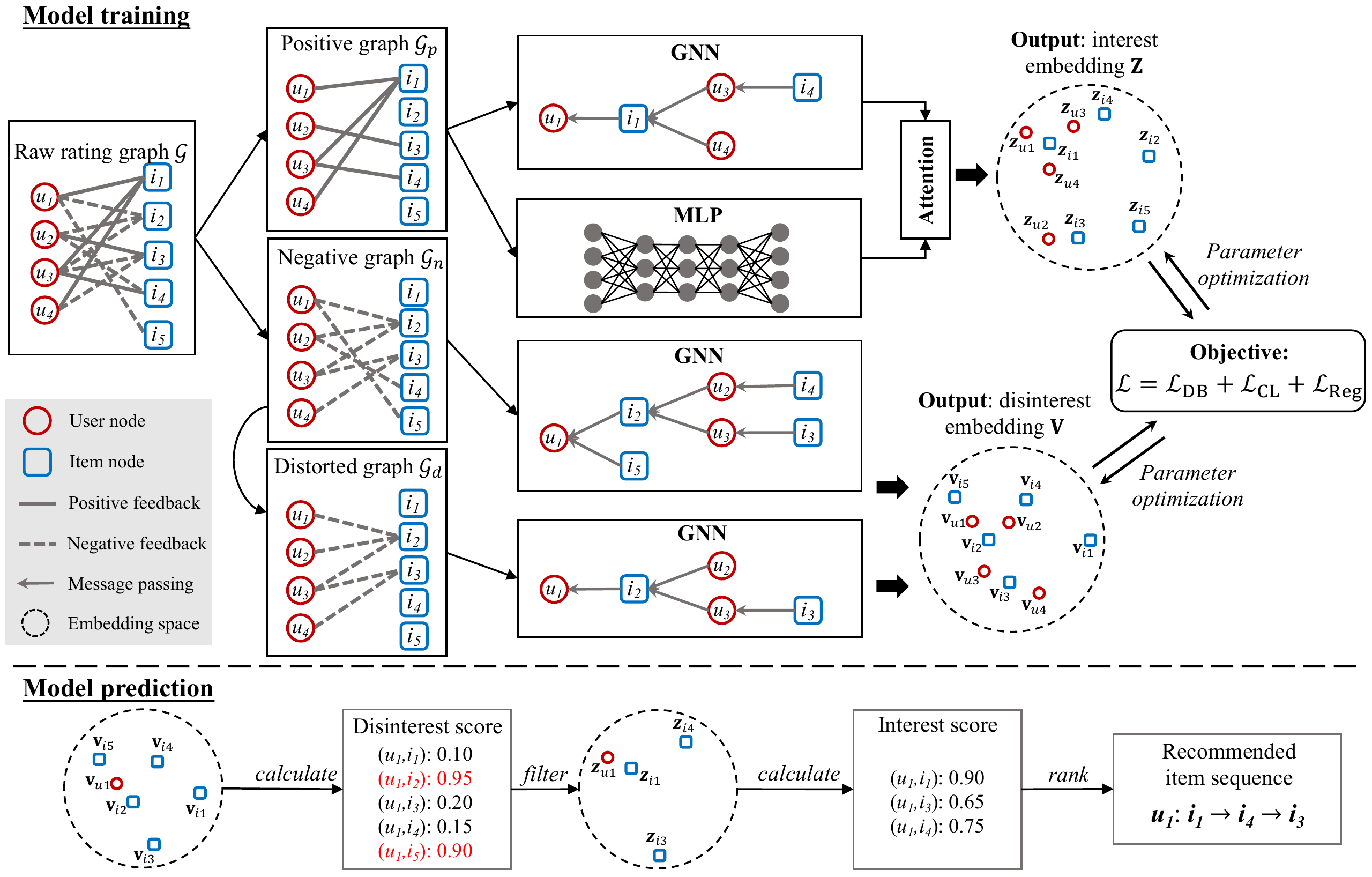}
\end{center}
\caption{The architecture of $\model$. In model training, $\model$ performs message passing on both $\mathcal{G}_{p}$ and $\mathcal{G}_{n}$ and contrastive learning on $\mathcal{G}_{n}$ to generate $\like$ embedding $\mathbf{Z}$ and $\dislike$ embedding $\mathbf{V}$. In model prediction, $\model$ recommends a sequence of items to each user based on a ranking method with a disinterest-score filter.}
\label{fig-model}
% \vspace*{-2mm}
\end{figure*}

\subsection{Model architecture}
\label{Model architecture}
The architecture of the $\model$ model is depicted in Figure~\ref{fig-model}. It consists of three key technical designs: message passing on the positive graph $\mathcal{G}{p}$ and the negative graph $\mathcal{G}{n}$, contrastive learning on the negative graph $\mathcal{G}_{n}$, and ranking with a disinterest-score filter.
In the message passing stage, information propagation takes place on both $\mathcal{G}{p}$ and $\mathcal{G}{n}$. This solution allows the model to leverage the structural information present in both graphs to enhance the representation learning process.
The contrastive learning stage focuses on the negative graph $\mathcal{G}_{n}$. By employing contrastive learning, the model denoises the negative feedback and generates robust embeddings that remain invariant to graph perturbations.
Finally, the ranking method with a disinterest-score filter is applied to generate the final recommendations. This method incorporates the learned embeddings from both the positive and negative graphs to rank the items and filter out items that do not align with the user's interests.

\subsubsection{Message passing on $\mathcal{G}_{p}$ and $\mathcal{G}_{n}$}
\label{Message passing on the positive graph and negative graph}
In contrast to prior work that primarily focuses on message passing on the positive graph $\mathcal{G}{p}$, $\model$ takes into account the high-order structural information in the negative graph $\mathcal{G}{n}$ as well.
In $\model$, we introduce two types of embeddings: $\like$ embeddings and $\dislike$ embeddings. These embeddings capture the relationships between liking and being liked, as well as disliking and being disliked, respectively, for each user or item.
To effectively aggregate and propagate these embeddings, $\model$ utilizes a technique called light graph convolution (LGC)~\cite{lightgcn}, which allows the embeddings to be updated and combined within the respective graph structures.
In the message passing process on the positive graph $\mathcal{G}_{p}$, the $\like$ embeddings $\mathbf{z}_{u}^{(k+1)}$ and $\mathbf{z}_{i}^{(k+1)}$ at the ($k{+}1$)-th layer are updated by summing the normalized $\like$ embeddings at the $k$-th layer:

\vspace*{-3mm}
\begin{equation}
\begin{aligned}
    \mathbf{z}_{u}^{(k+1)}&=\sum_{i\in \mathcal{N}_{p}(u)}\frac{1}{\sqrt{|\mathcal{N}_{p}(u)|}\sqrt{|\mathcal{N}_{p}(i)|}}\mathbf{z}_{i}^{(k)}, \\
    \mathbf{z}_{i}^{(k+1)}&=\sum_{u\in \mathcal{N}_{p}(i)}\frac{1}{\sqrt{|\mathcal{N}_{p}(i)|}\sqrt{|\mathcal{N}_{p}(u)|}}\mathbf{z}_{u}^{(k)}.
\end{aligned}
\label{message_passing_pos}
\end{equation}
The final $\like$ embeddings $\mathbf{z}_{u}$ and $\mathbf{z}_{i}$ can be obtained by averaging the $\like$ embeddings from all layers:

\vspace*{-3mm}
\begin{equation}
\begin{aligned}
    \mathbf{z}_{u}=\frac{1}{K+1}\sum_{k=0}^{K}\mathbf{z}_{u}^{(k)},\;\;\;\;
    \mathbf{z}_{i}=\frac{1}{K+1}\sum_{k=0}^{K}\mathbf{z}_{i}^{(k)},
\end{aligned}
\label{message_passing_pos_final}
\end{equation}
where $K$ is the total number of layers. In Eq.~(\ref{message_passing_pos_final}), $\mathbf{z}_{u}^{(0)}$ and $\mathbf{z}_{i}^{(0)}$ are trainable parameters that represent the initial embeddings for user $u$ and item $i$, respectively. These embeddings are randomly initialized before the model training process begins.
For the message passing process on the negative graph $\mathcal{G}_{n}$, the $\dislike$ embeddings $\mathbf{v}_{u}^{(k+1)}$ and $\mathbf{v}_{i}^{(k+1)}$ at the ($k+1$)-th layer are updated according to the following equations:

\vspace*{-3mm}
\begin{equation}
\begin{aligned}
    \mathbf{v}_{u}^{(k+1)}&=\sum_{i\in \mathcal{N}_{n}(u)}\frac{1}{\sqrt{|\mathcal{N}_{n}(u)|}\sqrt{|\mathcal{N}_{n}(i)|}}\mathbf{v}_{i}^{(k)}, \\
    \mathbf{v}_{i}^{(k+1)}&=\sum_{u\in \mathcal{N}_{n}(i)}\frac{1}{\sqrt{|\mathcal{N}_{n}(i)|}\sqrt{|\mathcal{N}_{n}(u)|}}\mathbf{v}_{u}^{(k)},
\end{aligned}
\label{message_passing_neg}
\end{equation}
The final $\dislike$ embeddings $\mathbf{v}_{u}$ and $\mathbf{v}_{i}$ are calculated by averaging the $\dislike$ embeddings of all layers:

\vspace*{-3mm}
\begin{equation}
\begin{aligned}
    \mathbf{v}_{u}=\frac{1}{K+1}\sum_{k=0}^{K}\mathbf{v}_{u}^{(k)},\;\;\;\;
    \mathbf{v}_{i}=\frac{1}{K+1}\sum_{k=0}^{K}\mathbf{v}_{i}^{(k)},
\end{aligned}
\label{message_passing_neg_final}
\end{equation}
where $\mathbf{v}_{u}^{(0)}$ and $\mathbf{v}_{i}^{(0)}$ are trainable parameters that are randomly initialized, similar to the initialization of $\like$ embeddings $\mathbf{z}_{u}^{(0)}$ and $\mathbf{z}_{i}^{(0)}$.
Correspondingly, the matrix forms of the above message-passing processes are as follows:

\vspace*{-3mm}
\begin{equation}
\begin{aligned}
    \mathbf{Z}'&{=}\frac{1}{K+1}\sum_{k=0}^{K}\mathbf{Z}^{(k)},\;\;\;\; 
    \mathbf{Z}^{(k+1)}{=}(\mathbf{D}_{p}^{-\frac{1}{2}}\mathbf{A}_{p}\mathbf{D}_{p}^{\frac{1}{2}})\mathbf{Z}^{(k)},
\label{Z'_calculation}
\end{aligned}
\end{equation}
\begin{equation}
\begin{aligned}
    \mathbf{V}&{=}\frac{1}{K+1}\sum_{k=0}^{K}\mathbf{V}^{(k)},\;\;\;\;
    \mathbf{V}^{(k+1)}{=}(\mathbf{D}_{n}^{-\frac{1}{2}}\mathbf{A}_{n}\mathbf{D}_{n}^{\frac{1}{2}})\mathbf{V}^{(k)},
\label{V_calculation}
\end{aligned}
\end{equation}
where $\mathbf{D}_{p}{=}\text{diag}(\mathbf{A}_{p}\mathbf{1}_{N{\times}N})$ and $\mathbf{D}_{n}{=}\text{diag}(\mathbf{A}_{n}\mathbf{1}_{N{\times}N})$ are the degree matrices of $\mathcal{G}_{p}$ and $\mathcal{G}_{n}$, respectively. Here $N{=}|\mathcal{U}{\cup}\mathcal{I}|$ is the number of all nodes in $\mathcal{G}$ and $\mathbf{1}_{N{\times}N}{\in}\mathbb{R}^{N{\times} N}$ is a square matrix of ones.

To incorporate dense non-graph information into the model, we use a two-layer MLP model to transform the initial $\like$ embeddings $\mathbf{Z}^{(0)}$ into a more expressive embedding $\mathbf{Z}''$:

\vspace*{-3mm}
\begin{equation}
\begin{aligned}
    \mathbf{Z}''&{=}\text{ReLU}(\text{ReLU}(\mathbf{Z}^{(0)}\mathbf{W}_{\text{MLP}}^{(1)})\mathbf{W}_{\text{MLP}}^{(2)}),
\label{Z_mlp}
\end{aligned}
\end{equation}
where $\mathbf{W}_{\text{MLP}}^{(1)},\mathbf{W}_{\text{MLP}}^{(2)}{\in} \mathbb{R}^{H\times H}$ are two trainable weight matrices to perform feature transformation.
Next, to determine the importance of the $\mathbf{Z}'$ and $\mathbf{Z}''$ embeddings in generating the final $\like$ embedding, we employ an attention mechanism. We introduce an attention layer that learns two importance scores $\alpha_{1},\alpha_{2}{\in}\mathbb{R}^{+}$ and yields the final $\like$ embedding $\mathbf{Z}$:

\vspace*{-3mm}
\begin{equation}
\begin{aligned}
    \hspace{-2mm}
    \mathbf{Z}{=}(\alpha_{1} &\mathbf{1}_{N\times H})\odot \mathbf{Z}' + (\alpha_{2} \mathbf{1}_{N\times H}) \odot \mathbf{Z}'', \\
    \hspace{-2mm}
    (\alpha_{1},\alpha_{2}){=}\text{Softmax}&(\text{Tanh}(\mathbf{Z}'\mathbf{W}_{\text{Att}}^{(1)})\mathbf{W}_{\text{Att}}^{(2)}, \text{Tanh}(\mathbf{Z}''\mathbf{W}_{\text{Att}}^{(1)})\mathbf{W}_{\text{Att}}^{(2)}),
\label{Z_calculation}
\end{aligned}
\end{equation}
where $\mathbf{W}_{\text{Att}}^{(1)}{\in} \mathbb{R}^{H{\times}H},\mathbf{W}_{\text{Att}}^{(2)}{\in}\mathbb{R}^{H{\times}1}$ are two trainable weight matrices and $\odot$ denotes the Hadamard product.

\subsubsection{Contrastive learning on $\mathcal{G}_{n}$}
\label{Contrastive learning on the positive graph}
Positive feedback serves as a reliable indicator of users' interests, while negative feedback is more susceptible to timeliness and contains more noise compared to positive feedback~\cite{negative_feature}. To address this issue, we propose a denoising approach in $\model$ by distorting the raw negative graph $\mathcal{G}_{n}$ into a new graph $\mathcal{G}_{d}$ and applying contrastive learning between the two graphs.
This approach is accomplished by applying edge removing, which is a widely used data augmentation strategy in graph contrastive learning, to the adjacency matrix $\mathbf{A}_{n}$ of the negative graph $\mathcal{G}_{n}$, resulting in the modified adjacency matrix $\mathbf{A}_{d}$:

\vspace*{-3mm}
\begin{equation}
\begin{aligned}
    \mathbf{A}_{d} = \mathbf{A}_{n}\odot \mathbf{P},\;\;\;\; \mathbf{P}\sim \mathcal{B}(1-p),
\label{distort_operation}
\end{aligned}
\end{equation}
where $\mathbf{P}$ is a random masking matrix drawn from a Bernoulli distribution with parameter $p$.
Then for the message passing process on $\mathcal{G}_{d}$, the $\dislike$ embeddings $\tilde{\mathbf{v}}_{u}^{(k+1)}$ and $\tilde{\mathbf{v}}_{i}^{(k+1)}$ at the ($k{+}1$)-th layer are updated using the following equations:

\vspace*{-3mm}
\begin{equation}
\begin{aligned}
    \tilde{\mathbf{v}}_{u}^{(k+1)}&=\sum_{i\in \mathcal{N}_{d}(u)}\frac{1}{\sqrt{|\mathcal{N}_{d}(u)|}\sqrt{|\mathcal{N}_{d}(i)|}}\tilde{\mathbf{v}}_{i}^{(k)}, \\
    \tilde{\mathbf{v}}_{i}^{(k+1)}&=\sum_{u\in \mathcal{N}_{d}(i)}\frac{1}{\sqrt{|\mathcal{N}_{d}(i)|}\sqrt{|\mathcal{N}_{d}(u)|}}\tilde{\mathbf{v}}_{u}^{(k)},
\end{aligned}
\end{equation}
where $\mathcal{N}_{d}(u){\subset}\mathcal{N}_{n}(u)$ and $\mathcal{N}_{d}(i){\subset}\mathcal{N}_{n}(i)$ are the neighbor sets of user $u$ and item $i$ in $\mathcal{G}_{d}$, respectively.
The final $\dislike$ embeddings $\tilde{\mathbf{v}}_{u}$ and $\tilde{\mathbf{v}}_{i}$ on $\mathcal{G}_{d}$ are calculated by averaging the $\dislike$ embeddings of all layers:

\vspace*{-3mm}
\begin{equation}
\begin{aligned}
    \tilde{\mathbf{v}}_{u}=\frac{1}{K+1}\sum_{k=0}^{K}\tilde{\mathbf{v}}_{u}^{(k)},\;\;\;\;
    \tilde{\mathbf{v}}_{i}=\frac{1}{K+1}\sum_{k=0}^{K}\tilde{\mathbf{v}}_{i}^{(k)},
\end{aligned}
\label{message_passing_distorted_final}
\end{equation}
where $\tilde{\mathbf{v}}_{u}^{(0)}{=}\mathbf{v}_{u}^{(0)}$ and $\tilde{\mathbf{v}}_{i}^{(0)}{=}\mathbf{v}_{i}^{(0)}$.
Correspondingly, the matrix form of the message-passing process on $\mathcal{G}_{d}$ is as follows:

\vspace*{-3mm}
\begin{equation}
\begin{aligned}
    \tilde{\mathbf{V}}&{=}\frac{1}{K+1}\sum_{k=0}^{K}\tilde{\mathbf{V}}^{(k)},\;\;\;\;
    \tilde{\mathbf{V}}^{(k+1)}{=}(\mathbf{D}_{d}^{-\frac{1}{2}}\mathbf{A}_{d}\mathbf{D}_{d}^{\frac{1}{2}})\tilde{\mathbf{V}}^{(k)},
\label{distort_V_calculation}
\end{aligned}
\end{equation}
where $\mathbf{D}_{d}{=}\text{diag}(\mathbf{A}_{d}\mathbf{1}_{N{\times}N})$ is the degree matrix of $\mathcal{G}_{d}$.

\subsubsection{Ranking with a disinterest-score filter}
\label{Ranking with a disinterest-score filter}
To calculate $\like$ scores, we utilize the matrix multiplication between the user embedding $\mathbf{z}_u$ and the item embedding $\mathbf{z}_i$, denoted as $S_{\text{it}}=\mathbf{z}_u\mathbf{z}_i^\text{T}$. This score represents the affinity between user $u$ and item $i$ based on their respective $\like$ embeddings.
Similarly, the $\dislike$ score is calculated as $S_{\text{dt}}=\mathbf{v}_u\mathbf{v}_i^\text{T}$. This score captures the disinterest or negative affinity between user $u$ and item $i$ based on their respective $\dislike$ embeddings.

The final recommended results for user $u$ are determined by applying a ranking function $\text{Rank}(\cdot)$ and a filtering function $\text{Filter}(\cdot)$ to the set of tuples $\left\{(u,i,S_{\text{it}},S_{\text{dt}})|i{\in} \mathcal{I}\right\}$.
The $\text{Filter}(\cdot)$ function returns a filtered set of tuples $\left\{(u,i,S_{\text{it}},S_{\text{dt}})|i{\in} \mathcal{I},S_{\text{dt}}{<}\delta\right\}$, ensuring that only items with $\dislike$ scores below the threshold $\delta$ are considered for recommendation.
The $\text{Rank}(\cdot)$ function ranks the filtered tuples based on $S_{\text{it}}$.
By combining these steps, the final recommended results for user $u$ can be expressed as:

\vspace*{-3mm}
\begin{equation}
\text{Result}(u){=}\text{Rank}(\text{Filter}(\left\{(u,i,S_{\text{it}},S_{\text{dt}})|i{\in} \mathcal{I}\right\},\delta)).
\end{equation}
This formulation allows us to generate recommendations that are ranked based on interest scores and filtered to exclude items with high disinterest scores.

% We calculate the $\like$ score with $S_{\text{it}}=\mathbf{z}_{u}\mathbf{z}_{i}^{\text{T}}$ and $\dislike$ score with $S_{\text{dt}}=\mathbf{v}_{u}\mathbf{v}_{i}^{\text{T}}$. The final recommended results of user $u$ are formulated as follows:

% \vspace*{-3mm}
% \begin{equation}
% \begin{aligned}
%     \text{Result}(u){=}\text{Rank}(\text{Filter}(\left\{(u,i,S_{\text{it}},S_{\text{dt}})|i{\in} \mathcal{I}\right\},\delta)),
% \end{aligned}
% \end{equation}
% where $\text{Rank}(\cdot)$ denotes the genral ranking function based on $S_\text{it}$ and $\text{Filter}(\cdot)$ denotes the filtering function based on $S_\text{dt}$. By defining a filtering threshold $\delta$, we filter those items whose $\dislike$ scores are higher than $\delta$:

% \vspace*{-3mm}
% \begin{equation}
% \begin{aligned}
%     \text{Filter}{:}\left\{(u,i,S_{\text{it}},S_{\text{dt}})|i{\in} \mathcal{I}\right\}\mapsto\left\{(u,i,S_{\text{it}},S_{\text{dt}})|i{\in} \mathcal{I},S_{\text{dt}}{>}\delta\right\}.
% \end{aligned}
% \end{equation}

\begin{algorithm}[t]
\algsetup{linenosize=\scriptsize}
\small
\caption{$\model$}%算法名字
\label{algorithm_model} 
\begin{algorithmic}[1] %这个1 表示每一行都显示数字
\REQUIRE 
    Positive graph $\mathcal{G}_{p}$, negative graph $\mathcal{G}_{n}$, trainable parameters $\Theta_{\text{Emb}}{=}\left\{\mathbf{Z}^{(0)},\mathbf{V}^{(0)}\right\}$ and $\Theta_{\text{NN}}{=}\left\{ \mathbf{W}_{\text{MLP}}^{(1)},\mathbf{W}_{\text{MLP}}^{(2)},\mathbf{W}_{\text{Att}}^{(1)},\mathbf{W}_{\text{Att}}^{(2)} \right\}$, embedding size $H$, GNNs layer number $K$, hyperparameters $p,b,\delta,\lambda_{1},\lambda_{2},\tau$.
\ENSURE
    Interest embedding matrix $\mathbf{Z}$, disinterest embedding matrix $\mathbf{V}$. \\
\STATE Initialize $\Theta_{\text{Emb}}$ and $\Theta_{\text{NN}}$ via the Glorot method;  \\
\STATE Initialize embedding matrices: $\mathbf{Z}\leftarrow \mathbf{Z}^{(0)}$, $\mathbf{V}\leftarrow \mathbf{V}^{(0)}$,
$\tilde{\mathbf{V}}\leftarrow \mathbf{V}^{(0)}$;  \\
\STATE Distort $\mathcal{G}_{n}$ into $\mathcal{G}_{d}$ according to Eq.~(\ref{distort_operation}); \\
\WHILE{not converged}
    \STATE Generate training set $\mathcal{D}_{p}$ from $\mathcal{G}_{p}$ based on Eq.~(\ref{training_set_1}); \\
    \STATE Generate training set $\mathcal{D}_{n}$ from $\mathcal{G}_{n}$ based on Eq.~(\ref{training_set_2}); \\
    \FOR{each mini-batch $\mathcal{B}_{p}{\subset} \mathcal{D}_{p}$}
        \STATE Calculate $\mathbf{Z}'$ according to Eq.~(\ref{Z'_calculation}); \\
        \STATE Calculate $\mathbf{Z}''$ according to Eq.~(\ref{Z_mlp}); \\
        \STATE Update $\mathbf{Z}$ according to Eq.~(\ref{Z_calculation}); \\
    \ENDFOR
    \FOR{each mini-batch $\mathcal{B}_{n}{\subset} \mathcal{D}_{n}$}
        \STATE Update $\mathbf{V}$ according to Eq.~(\ref{V_calculation}); \\
        \STATE Update $\tilde{\mathbf{V}}$ according to Eq.~(\ref{distort_V_calculation});
    \ENDFOR
    \STATE Calculate $\mathcal{L}_{\text{DB}}$ according to Eq.~(\ref{loss_db}); \\
    \STATE Calculate $\mathcal{L}_{\text{CL}}$ according to Eq.~(\ref{loss_cl}); \\
    \STATE $\mathcal{L}_{\text{Reg}}\leftarrow \left \| \Theta_{\text{Emb}} \right \|^{2}$; \\
    \STATE $\mathcal{L}\leftarrow \mathcal{L}_{\text{DB}} + \lambda_{1}{\cdot}\mathcal{L}_{\text{CL}} + \lambda_{2}{\cdot} \mathcal{L}_{\text{Reg}}$ \\
    Update $\Theta_{\text{Emb}}$ and $\Theta_{\text{NN}}$ by taking one step of gradient descent on $\mathcal{L}$;
\ENDWHILE
\RETURN $\mathbf{Z}$, $\mathbf{V}$.
\end{algorithmic}
\end{algorithm}
\vspace*{-3mm}

\subsection{Optimization}
\label{Optimization}
To construct the training sets for the positive graph $\mathcal{G}_{p}$ and the negative graph $\mathcal{G}_{n}$, we define two sets of training examples $\mathcal{D}_{p}$ and $\mathcal{D}_{n}$ as follows:

\vspace*{-3mm}
\begin{equation}
\begin{aligned}
    \mathcal{D}_{p}=\left \{ (u,i,j)|(u,i){\in} \mathcal{E}_{p},j{\notin} \mathcal{N}_{p}(u) \right \},
\label{training_set_1}
\end{aligned}
\end{equation}

\vspace*{-3mm}
\begin{equation}
\begin{aligned}
    \mathcal{D}_{n}=\left \{ (u,i,j)|(u,i){\in} \mathcal{E}_{n},j{\notin} \mathcal{N}_{n}(u) \right \}.
\label{training_set_2}
\end{aligned}
\end{equation}
Furthermore, we leverage mini-batch learning to train $\model$, then each mini-batch on $\mathcal{G}_{p}$ and $\mathcal{G}_{n}$ are denoted as $\mathcal{B}_{p}{\subset} \mathcal{D}_{p}$ and $\mathcal{B}_{n}{\subset} \mathcal{D}_{n}$, respectively.

The trainable parameter group of $\model$ consists of two parts: the embeddings $\Theta_{\text{Emb}}{=}\left\{ \mathbf{Z}^{(0)}, \mathbf{V}^{(0)} \right\}$ of the 0-th layer, and the neural network parameters $\Theta_{\text{NN}}{=}\left\{ \mathbf{W}_{\text{MLP}}^{(1)}, \mathbf{W}_{\text{MLP}}^{(2)}, \mathbf{W}_{\text{Att}}^{(1)}, \mathbf{W}_{\text{Att}}^{(2)} \right\}$, which include the weight matrices for the MLP layers and attention layers. The overall loss function $\mathcal{L}$ is defined as follows:

\vspace*{-3mm}
\begin{equation}
\begin{aligned}
    \mathcal{L} = \mathcal{L}_{\text{DB}} + \lambda_{1}{\cdot}\mathcal{L}_{\text{CL}} + \lambda_{2}{\cdot} \mathcal{L}_{\text{Reg}},
\label{total_loss}
\end{aligned}
\end{equation}
where $\mathcal{L}_{\text{Reg}}{=}\left \| \Theta_{\text{Emb}} \right \|^{2}$ denotes the L2 regularization term of the 0-th layer embeddings. $\lambda_{1}$ and $\lambda_{2}$ are two hyperparameters that control the strength of contrastive learning and L2 regularization, respectively.
In order to incorporate the feedback information from both $\mathcal{G}_{p}$ and $\mathcal{G}_{n}$, we propose a dual feedback-aware BPR loss $\mathcal{L}_{\text{DB}}$ inspired by the Bayesian personalized ranking (BPR) loss~\cite{bpr_loss}:

\vspace*{-3mm}
\begin{equation}
\begin{aligned}
    \hspace{-2mm}
    \mathcal{L}_{\text{DB}} &= {-}\sum_{(u,i,j){\in} \mathcal{B}_{p}}\text{ln}\sigma(\hat{y}_{u,i}{-}\hat{y}_{u,j}){-}\sum_{(u,i,j){\in} \mathcal{B}_{n}}\text{ln}\sigma(\hat{y}_{u,j}{-}\hat{y}_{u,i}), \\
    \hspace{-2mm}
    \hat{y}_{u,i}&{=}\left\{\begin{matrix}
    \;\;\; \mathbf{z}_{u}\mathbf{z}_{i}^{\text{T}},\; \text{if}\;  (u,i,j){\in} \mathcal{B}_{p} \\ 
    b{\cdot} \mathbf{v}_{u}\mathbf{v}_{i}^{\text{T}},\; \text{if}\;  (u,i,j){\in} \mathcal{B}_{n}
    \end{matrix}\right. \;\;
    \hat{y}_{u,j}{=}\left\{\begin{matrix}
    \; \mathbf{z}_{u}\mathbf{z}_{j}^{\text{T}},\; \text{if}\;  (u,i,j){\in} \mathcal{B}_{p} \\
    \; \mathbf{v}_{u}\mathbf{v}_{j}^{\text{T}},\; \text{if}\;  (u,i,j){\in} \mathcal{B}_{n}
    \end{matrix}\right.
\label{loss_db}
\end{aligned}
\end{equation}
where $\sigma(x){=}\frac{1}{1+\text{exp}(-x)}$ is the Sigmoid function and $b{>}1$ is a feedback-aware coefficient.
The presence of $b$ ensures the following priority order: positive feedback > negative feedback > no feedback. This priority implies that positive feedback is given higher importance than negative feedback, and both positive and negative feedback are considered more valuable than no feedback.
In addition, we design the contrastive objective $\mathcal{L}_{\text{CL}}$ on $\mathcal{G}_{n}$ via the InfoNCE loss~\cite{infonce}:

\vspace*{-3mm}
\begin{equation}
\begin{aligned}
    \hspace{-2mm}
    \mathcal{L}_{\text{CL}} {=} {-}\sum_{u{\in} \mathcal{U}} \text{ln}\frac{\text{exp}(\frac{\mathbf{v}_{u}\tilde{\mathbf{v}}_{u}^{\text{T}}}{\tau})}{\sum_{u'{\in} \mathcal{U}}\text{exp}(\frac{\mathbf{v}_{u}\tilde{\mathbf{v}}_{u'}^{\text{T}}}{\tau})}
    {-}\sum_{i{\in} \mathcal{I}} \text{ln}\frac{\text{exp}(\frac{\mathbf{v}_{i}\tilde{\mathbf{v}}_{i}^{\text{T}}}{\tau})}{\sum_{i'{\in} \mathcal{I}}\text{exp}(\frac{\mathbf{v}_{i}\tilde{\mathbf{v}}_{i'}^{\text{T}}}{\tau})},
\label{loss_cl}
\end{aligned}
\end{equation}
where $\tau$ is a temperature coefficient.
This objective allows us to leverage the contrastive learning framework to enhance the robustness and discriminative power of $\dislike$ embeddings in the recommendation process.
The complete procedure of $\model$ is summarized in Algorithm~\ref{algorithm_model}.

\section{Experiment}
\label{Experiment}
In this section, we provide descriptions of the four real-world datasets (Section~\ref{Datasets}) and five baselines (Section~\ref{Baselines}) used in our experiments. We also introduce the metrics (Section~\ref{Metrics}) and hyperparameter setups (Section~\ref{Hyperparameter Setups}). Furthermore, we compare the performance of different methods and conduct a comprehensive evaluation of the performance of $\model$ (Section~\ref{Results}).

\subsection{Datasets}
\label{Datasets}
We evaluate our approach on four real-world datasets: MovieLens-1M (ML-1M), Amazon-Book, Yelp, and KuaiRec.

\begin{itemize}[leftmargin=*]
    \item \textbf{ML-1M} (\href{http://q6e9.cn/VMNQw}{\textcolor{black}{http://q6e9.cn/VMNQw}}): This widely-used movie review dataset consists of approximately 6,000 users and 4,000 movies. Users rate movies on a 5-star scale, and each user has provided at least 20 ratings.
    \item \textbf{Amazon-Book} (\href{https://61a.life/K7oer}{\textcolor{black}{https://61a.life/K7oer}}): We selected the Amazon-Book dataset from a large crawl of product reviews on Amazon. The dataset comprises around 35,000 users, 38,000 items, and 1.9 million 5-star ratings. Similar to previous work~\cite{vbpr,siren}, we removed users or items with fewer than 20 interactions.
    \item \textbf{Yelp} (\href{https://x064.cn/Jak1U}{\textcolor{black}{https://x064.cn/Jak1U}}): This dataset consists of reviews for local businesses. It includes approximately 41,000 users, 30,000 businesses, and 2.1 million 5-star ratings. Like the Amazon-Book dataset, we excluded users or businesses with fewer than 20 interactions.
    \item \textbf{KuaiRec} (\href{https://54z.life/DuQDC}{\textcolor{black}{https://54z.life/DuQDC}}): This real-world dataset was collected from the recommendation logs of Kuaishou, a video-sharing mobile app. It contains around 7,100 users, 10,000 short videos (each with multiple tags), and a user-video interaction matrix.
\end{itemize}

For ML-1M, Amazon-Book, and Yelp, we use the threshold of 3.5 to split the original ratings as binary signals. For KuaiRec, as suggested by the authors in~\cite{kuairec}, we use the rule of ``whether the video watch ratio is higher than 2.0'' to achieve binary signals.
The detailed statistics of the above four datasets are shown in Table~\ref{Statistics of four real-world datasets}. In the training set of KuaiRec, the number of negative ratings is far higher than that of positive ratings, which provides a more realistic and biased training environment compared to the other three datasets.

\subsection{Baselines}
\label{Baselines}
We compare $\model$ with five state-of-the-art GNN-based recommendation models.

\begin{itemize}[leftmargin=*]
    \item \textbf{NGCF}~\cite{NGCF}: NGCF is a GNN-based recommendation framework that explicitly incorporates high-order collaborative signals from the user-item bipartite graph through embedding propagation.
    \item \textbf{LR-GCCF}~\cite{LR-GCCF}: LR-GCCF incorporates the GCN model into the recommender system. Instead of employing non-linear transformations in the GCN, LR-GCCF utilizes linear embedding propagations. Additionally, it introduces a residual network structure to address the over-smoothing issue that can arise from applying multiple layers of graph convolutions.
    \item \textbf{LightGCN}~\cite{lightgcn}: LightGCN redesigns a light graph convolution structure specific to recommendations by abandoning the use of feature transformation and nonlinear activation. This approach aims to simplify the model while maintaining competitive performance.
    \item \textbf{SGCN}~\cite{SGCN}: SGCN leverages balance theory to aggregate and propagate information in a signed graph. By considering balanced and unbalanced embeddings, SGCN effectively captures the information from both positive and negative feedback signals.
    \item \textbf{SiReN}~\cite{siren}: SiReN is designed for signed bipartite graphs. It utilizes a GNN model and an MLP model to generate two sets of embeddings for the partitioned graph. Additionally, SiReN designs a sign-aware BPR loss to differentiate the effects of high-rating and low-rating items.
\end{itemize}

\begin{table}[t]
\setlength{\abovecaptionskip}{-0mm} 
\setlength{\belowcaptionskip}{-0mm}
\setlength\tabcolsep{5.0pt}
\small
\caption{Statistics of four real-world datasets. ``Ratio'' denotes the number ratio between positive and negative ratings in the training set.}
\label{Statistics of four real-world datasets}
\begin{center}
\begin{threeparttable}
\begin{tabular}{ccccccc}
\toprule[1pt]
\bf Dataset &\bf \#User  &\bf \#Item  &\bf \#Rating  &\bf Density (\%) &\bf Ratio \\
\midrule[0.5pt]
ML-1M &6,040 &3,952 &1,000,209 &4.19 &1:0.73 \\
Amazon-Book &35,736 &38,121 &1,960,674 &0.14 &1:0.24 \\
Yelp &41,772 &30,037 &2,116,215 &0.16 &1:0.47 \\
KuaiRec &7,176 &10,728 &761,425 &0.98 &1:13.30 \\
% ML-1M: 459962,340205
% Amazon: 1252292,302056
% Yelp: 1141287, 535055
% KuaiRec: 52789,702096
\bottomrule[1pt]
\end{tabular}
\end{threeparttable}
\end{center}
\vspace*{-6mm}
\end{table}

\begin{table*}[t]
\setlength{\abovecaptionskip}{-0mm} 
\setlength{\belowcaptionskip}{-0mm}
\setlength\tabcolsep{2.0pt}
\small
\caption{Results (\%) of baselines and our method on ML-1M, Amazon-Book, Yelp, and KuaiRec. In each column, the best result is bolded. The results of the methods marked with ``$^{\dagger}$'' are from~\cite{siren}. For the methods marked with ``$^{\star}$'', we run each of them five times with default hyperparameter settings.}
\label{result-three-datasets}
\begin{center}
\begin{threeparttable}
\begin{tabular}{ccccc|ccc|ccc}
\toprule[1pt]
\multirow{2}*{\bf Dataset} &\multirow{2}*{\bf Method} &\multicolumn{3}{c|}{$K = 5$} &\multicolumn{3}{c|}{$K = 10$} &\multicolumn{3}{c}{$K = 15$}  \\
& &$Precision@K$ &$Recall@K$ &$nDCG@K$ &$Precision@K$ &$Recall@K$ &$nDCG@K$ &$Precision@K$ &$Recall@K$ &$nDCG@K$ \\
\midrule[1.0pt]
% \multirow{8}*{\rotatebox{90}{ML-1M}} &BPRMF &23.60±0.58 &7.46±0.17 &25.36±0.81 &19.99±0.44 &12.27±0.19 &23.63±0.63 &17.72±0.33 &16.08±0.16 &23.14±0.53 \\
% &NeuMF &27.85±0.54 &9.66±0.11 &29.90±0.35 &23.97±0.30 &15.99±0.11 &28.47±0.35 &21.38±0.27 &20.98±0.23 &28.27±0.32 \\
\multirow{6}*{\rotatebox{90}{ML-1M}} &NGCF$^{\dagger}$ &29.73±0.43 &10.99±0.26 &32.38±0.45 &24.77±0.23 &17.48±0.25 &30.31±0.33 &21.74±0.22 &22.29±0.27 &29.85±0.29  \\
&LR-GCCF$^{\dagger}$ &30.52±0.33 &11.40±0.23 &33.30±0.44 &25.39±0.27 &18.02±0.31 &31.17±0.39 &22.20±0.25 &22.92±0.46 &30.66±0.42 \\
&LightGCN$^{\dagger}$ &32.18±0.22 &12.06±0.11 &35.19±0.23 &26.79±0.13 &19.09±0.16 &32.97±0.18 &23.49±0.16 &24.32±0.29 &32.49±0.22 \\
&SGCN$^{\dagger}$ &24.84±033 &9.10±0.17 &26.83±0.35 &18.73±0.20 &14.92±0.26 &25.47±0.24 &18.73±0.20 &19.32±0.37 &25.30±0.26\\
&SiReN$^{\dagger}$ &33.28±0.54  &12.79±0.27  &36.37±0.55  &27.74±0.37  &20.16±0.33  &34.23±0.47  &24.44±0.25  &25.69±0.29  &33.88±0.40\\
&$\model$ (Ours)$^{\star}$ &\textbf{33.66±0.14}  &\textbf{13.26±0.17}  &\textbf{36.90±0.25}  &\textbf{27.97±0.14}  &\textbf{20.50±0.18}  &\textbf{34.70±0.13}  &\textbf{24.66±0.22} &\textbf{25.95±0.09} &\textbf{34.37±0.14} \\
% &Ours &\textbf{36.66±0.13}  &\textbf{14.10±0.15}  &\textbf{40.01±0.22} &\textbf{30.55±0.24}   &\textbf{21.83±0.17}  &\textbf{37.60±0.08}  &\textbf{26.53±0.10} &\textbf{27.29±0.17} &\textbf{36.84±0.22} \\
% &SiReN (reprod) &32.25±0.26 &12.67±0.14  &35.24±0.11  &27.04±0.13 &19.86±0.10  &33.37±0.19  &23.93±0.12 &25.28±0.20 &33.16±0.12 \\
\midrule[1.0pt]
% \multirow{8}*{\rotatebox{90}{Amazon-Book}} &BPRMF &2.98±0.39  &2.09±0.35  &3.31±0.46  &2.63±0.33  &3.65±0.57  &3.71±0.53  &2.43±0.29  &5.00±0.74  &4.19±0.58\\
% &NeuMF &4.02±0.21 &2.71±0.07 &4.48±0.20 &3.39±0.18 &4.52±0.10 &4.83±0.18 &3.03±0.14 &5.97±0.12 &5.30±0.17\\
\multirow{6}*{\rotatebox{90}{Amazon-Book}} &NGCF$^{\dagger}$ &4.63±0.14 &3.20±0.07 &5.18±0.11 &3.91±0.14 &5.32±0.08 &5.62±0.10 &4.03±1.27 &7.06±0.07 &6.18±0.09\\
&LR-GCCF$^{\dagger}$ &4.69±0.16 &3.24±0.02 &5.27±0.12 &3.99±0.14 &5.44±0.05 &5.74±0.09 &3.57±0.13 &7.21±0.04 &6.31±0.08\\
&LightGCN$^{\dagger}$ &5.29±0.15 &3.62±0.07 &5.96±0.11 &4.43±0.13 &5.95±0.08 &6.38±0.07 &3.93±0.11 &7.81±0.08 &6.98±0.07\\
&SGCN$^{\dagger}$ &3.90±0.23 &2.67±0.12 &4.33±0.24 &3.36±0.18 &4.54±0.18 &4.75±0.23 &3.04±0.16 &6.09±0.21 &5.27±0.24\\
&SiReN$^{\dagger}$ &6.78±0.25 &4.74±0.05 &7.66±0.02 &5.65±0.25 &7.75±0.08 &8.23±0.13 &4.97±0.18 &10.09±0.11 &8.97±0.11\\
&$\model$ (Ours)$^{\star}$ &\textbf{7.31±0.05}  &\textbf{4.95±0.19}  &\textbf{8.14±0.08}   &\textbf{6.07±0.22}  &\textbf{8.05±0.18}  &\textbf{8.69±0.26}  &\textbf{5.32±0.20} &\textbf{10.41±0.05} &\textbf{9.45±0.11} \\
\midrule[1.0pt]
% \multirow{8}*{\rotatebox{90}{Yelp}} &BPRMF &1.24±0.13 &0.96±0.09 &1.37±0.13 &1.16±0.14 &1.80±0.20 &1.65±0.17 &1.11±0.13 &2.57±0.26 &1.94±0.21\\
% &NeuMF &1.99±0.11 &1.50±0.12 &2.26±0.15 &1.74±0.09 &2.62±0.17 &2.54±0.15 &1.59±0.07 &3.58±0.19 &2.88±0.16\\
\multirow{6}*{\rotatebox{90}{Yelp}} &NGCF$^{\dagger}$ &2.85±0.12 &2.26±0.07 &3.29±0.10 &2.43±0.09 &3.83±0.10 &3.68±0.10 &2.19±0.07 &5.15±0.08 &4.13±0.09\\
&LR-GCCF$^{\dagger}$ &3.03±0.14 &2.40±0.06 &3.51±0.13 &2.58±0.10 &4.05±0.09 &3.92±0.11 &2.32±0.08 &5.43±0.10 &4.39±0.12\\
&LightGCN$^{\dagger}$ &3.33±0.11 &2.59±0.04 &3.86±0.09 &2.81±0.09 &4.35±0.07 &4.27±0.08 &2.51±0.08 &5.82±0.10 &4.76±0.08\\
&SGCN$^{\dagger}$ &2.93±0.10 &2.26±0.06 &3.32±0.10 &2.56±0.08 &3.95±0.11 &3.77±0.10 &2.32±0.07 &5.38±0.19 &4.26±0.12\\
&SiReN$^{\dagger}$ &4.20±0.09 &3.32±0.05 &4.88±0.07 &3.52±0.07 &5.54±0.11 &5.39±0.07 &3.14±0.06 &7.37±0.12 &6.00±0.06\\
&$\model$ (Ours)$^{\star}$ &\textbf{4.75±0.11}  &\textbf{3.56±0.06} &\textbf{5.49±0.11} &\textbf{3.93±0.13}  &\textbf{5.89±0.07} &\textbf{5.94±0.14} &\textbf{3.51±0.09} &\textbf{7.83±0.05} &\textbf{6.59±0.10} \\
\midrule[1.0pt]
\multirow{6}*{\rotatebox{90}{KuaiRec}} &NGCF$^{\star}$ &2.05±0.18  &5.05±0.20  &3.85±0.09  &1.88±0.04  &8.39±0.17 &5.10±0.18  &1.69±0.21 &10.47±0.14 &5.80±0.14 \\
&LR-GCCF$^{\star}$ &4.84±0.21 &11.39±0.25 &9.37±0.17 &3.60±0.09 &15.43±0.11 &10.79±0.14 &2.96±0.09 &17.80±0.18 &11.60±0.22 \\
% &SGCN &16.69±0.12  &13.94±0.10  &17.26±0.24  &14.19±0.25  &20.72±0.08  &18.15±0.19  &13.35±0.17 &27.88±0.11 &20.28±0.08 \\
&LightGCN$^{\star}$ &23.83±0.19  &33.76±0.22  &39.11±0.07   &17.58±0.05  &43.16±0.14  &41.04±0.20  &14.84±0.21 &50.95±0.12 &43.65±0.15 \\
&SGCN$^{\star}$ &0.16±0.01  &0.13±0.00  &0.17±0.00  &0.14±0.01  &0.20±0.00  &0.18±0.01  &0.13±0.02 &0.27±0.00 &0.20±0.01 \\
&SiReN$^{\star}$ &24.50±0.08  &33.33±0.10  &40.07±0.08 &17.73±0.22  &42.75±0.17 &41.44±0.13  &14.81±0.16 &49.90±0.05 &43.72±0.09 \\
&$\model$ (Ours)$^{\star}$ &\textbf{25.85±0.10}  &\textbf{34.61±0.11}  &\textbf{41.91±0.20}  &\textbf{19.39±0.17}  &\textbf{46.03±0.07}  &\textbf{44.13±0.05}  &\textbf{16.16±0.16} &\textbf{53.47±0.15} &\textbf{46.55±0.10} \\
\bottomrule[1pt]
\end{tabular}
\end{threeparttable}
\end{center}
% \vspace*{-3mm}
\end{table*}

\subsection{Metrics}
\label{Metrics}
We evaluate the effectiveness of $\model$ using three performance metrics: $Precision@K$, $Recall@K$, and $nDCG@K$ (normalized discounted cumulative gain$@K$). These metrics provide insights into the accuracy, completeness, and ranking quality of the recommendation results.
$Precision@K$ measures the proportion of relevant items among the top-$K$ recommended results for a user:

\vspace*{-3mm}
\begin{equation}
    Precision@K = \frac{1}{|\mathcal{U}|}\sum_{u\in \mathcal{U}} \frac{|GT_{u}\cap R_{u}(K)|}{K},
\end{equation}
where $GT_{u}$ denotes the ground truth item set liked by user $u$ in the test set and $R_{u}(K)$ denotes the recommended top-$K$ items for user $u$.
$Recall@K$ quantifies the proportion of relevant items among all correct results for a user:

\vspace*{-3mm}
\begin{equation}
    Recall@K = \frac{1}{|\mathcal{U}|}\sum_{u\in \mathcal{U}} \frac{|GT_{u}\cap R_{u}(K)|}{|GT_{u}|}.
\end{equation}
$nDCG@K$ is a ranking quality measurement that assigns higher values to relevant items appearing at higher ranks:

\vspace*{-3mm}
\begin{equation}
\begin{aligned}
    nDCG@K {=} \frac{1}{|\mathcal{U}|}&\sum_{u\in \mathcal{U}} \frac{DCG_{u}@K}{IDCG_{u}@K}, \\
    DCG_{u}@K {=} \sum_{i=1}^{K}\frac{G_{u}(i)}{log_{2}(i+1)}&,\;\;
    iDCG_{u}@K {=} \sum_{i=1}^{K}\frac{1}{log_{2}(i+1)},
\end{aligned}
\end{equation}
where $G_{u}(i)$ equals 1 if the item at rank $i$ in the recommended list is in the ground truth item set $GT_{u}$, and 0 otherwise.

\subsection{Hyperparameter Setups}
\label{Hyperparameter Setups}
In the experiments, we set the embedding size of $\model$ to 64, similar to LightGCN and SiReN. The embedding parameters of $\model$ are initialized using the Glorot method~\cite{Glorot}.
We use the Adam optimizer~\cite{adam} with a default learning rate of 5e-3 to optimize $\model$. The training process of $\model$ employs mini-batch learning, where the default batch size is set to 1,024. We train $\model$ for a total of 1,000 epochs for all datasets.
$\model$ incorporates L2 regularization with a coefficient of 0.01 on KuaiRec and 0.05 on the other three datasets. Negative sampling is employed during training, and the number of negative samples is set to 1 on KuaiRec and 40 on the other three datasets.
The architecture of $\model$ consists of 4 layers of GNNs and 2 layers of MLP in total. The temperature value used in the contrastive loss is set to 0.8. Additionally, the dropout rate for the MLP layer or attention layer is set to 0.5. The filter in $\model$ utilizes a $\dislike$ score threshold of 0.5 by default.
The implementation of $\model$ is done using PyTorch. The source code is available at \href{https://reurl.cc/0ELqO6}{\textcolor{black}{https://reurl.cc/0ELqO6}}.

For ML-1M, Amazon-Book, and Yelp datasets, we perform 5-fold cross-validation by splitting each dataset into training and test sets. The training set contains 80\% of the ratings, while the remaining 20\% constitutes the test set. As for KuaiRec, following the suggestion in the original paper~\cite{kuairec}, we use the user-item interactions from the fully-observed small matrix as the test set, and the remaining interactions are used for training.

\begin{table*}[t]
\setlength{\abovecaptionskip}{-0mm} 
\setlength{\belowcaptionskip}{-0mm}
\setlength\tabcolsep{1.0pt}
\small
\caption{Results (\%) of ablation studies on ML-1M and KuaiRec. Here ``MP'', ``GCL'', and ``Filter'' denote message passing, graph contrastive learning, and the disinterest-score filter, respectively.}
\label{ablation-study}
\begin{center}
\begin{threeparttable}
\begin{tabular}{cccccc|ccc|ccc}
\toprule[1pt]
\multirow{2}*{\bf Dataset} &\multirow{2}*{\bf Variant} &\multirow{2}*{\bf Description} &\multicolumn{3}{c|}{$K = 5$} &\multicolumn{3}{c|}{$K = 10$} &\multicolumn{3}{c}{$K = 15$}  \\
& & &$Precision@K$ &$Recall@K$ &$nDCG@K$ &$Precision@K$ &$Recall@K$ &$nDCG@K$ &$Precision@K$ &$Recall@K$ &$nDCG@K$ \\
\midrule[1.0pt]
\multirow{5}*{\rotatebox{90}{ML-1M}} &A &MP on $\mathcal{G}_{n}$ &0.64±0.02  &0.13±0.04  &0.68±0.03  &0.62±0.01  &0.26±0.01  &0.67±0.02  &0.61±0.02 &0.43±0.01 &0.70±0.03 \\
&B &MP on $\mathcal{G}_{p}$ &31.51±0.15  &12.11±0.10  &34.49±0.20  &26.35±0.22  &19.23±0.11  &32.59±0.17   &23.32±0.20 &24.47±0.19 &32.34±0.15 \\
&C &MP on $\mathcal{G}_{p}$ \& $\mathcal{G}_{n}$ &32.65±0.08  &12.87±0.18  &35.92±0.15  &27.49±0.23  &20.35±0.09  &34.13±0.11  &24.17±0.14 &25.67±0.13 &33.79±0.13 \\
&D &Variant-C + GCL &33.46±0.11  &13.05±0.15   &36.56±0.21  &27.77±0.20  &20.40±0.11  &34.45±0.09  &24.36±0.12 &25.70±0.17 &34.05±0.04 \\
% &E &$\mathrm{D_{+Filter}}$ &36.66±0.13  &14.10±0.15  &40.01±0.22 &30.55±0.24   &21.83±0.17  &37.60±0.08  &26.53±0.10 &27.29±0.17 &36.84±0.22 \\
&$\model$ &Variant-D + Filter &33.66±0.14  &13.26±0.17  &36.90±0.25  &27.97±0.14  &20.50±0.18  &34.70±0.13  &24.66±0.22 &25.95±0.09 &34.37±0.14 \\
\midrule[1.0pt]
\multirow{5}*{\rotatebox{90}{KuaiRec}} &A &MP on $\mathcal{G}_{n}$ &5.54±0.00  &5.13±0.01  &6.54±0.01  &5.40±0.01  &10.29±0.00  &8.09±0.02  &5.61±0.01 &15.78±0.01 &10.08±0.00 \\
&B &MP on $\mathcal{G}_{p}$ &24.22±0.10  &33.25±0.13  &40.05±0.13  &17.61±0.08  &42.19±0.23 &41.60±0.16 &14.64±0.15 &49.36±0.17 &43.83±0.12 \\
&C &MP on $\mathcal{G}_{p}$ \& $\mathcal{G}_{n}$ &24.70±0.14  &32.90±0.09 &40.52±0.20   &17.70±0.22  &42.67±0.27  &41.59±0.09  &14.92±0.14 &50.12±0.18 &44.03±0.13 \\
&D &Variant-C + GCL &24.94±0.18  &34.04±0.22     &41.21±0.06  &19.37±0.14  &45.81±0.04  &44.10±0.28  &15.88±0.19 &53.36±0.06 &46.28±0.18 \\
&$\model$ &Variant-D + Filter &25.85±0.10  &34.61±0.11  &41.91±0.20  &19.39±0.17  &46.03±0.07  &44.13±0.05  &16.16±0.16 &53.47±0.15 &46.55±0.10 \\
\bottomrule[1pt]
\end{tabular}
\end{threeparttable}
\end{center}
% \vspace*{-3mm}
\end{table*}

\subsection{Experimental Results}
\label{Results}
We conduct experiments to answer the following four key research questions:

\begin{itemize}[leftmargin=*]
    \item \textbf{RQ1:} Does $\model$ improve overall recommendation performance compared to other GNN-based methods (Section~\ref{Comparison of overall performance})?
    \item \textbf{RQ2:} How do different components in $\model$ affect its performance (Section~\ref{Ablation studies})?
    % \item \textbf{RQ3:} What is the optimal approach for utilizing negative feedback effectively (Section~\ref{Comparison with multi-behavior model})?
    \item \textbf{RQ3:} How robust is $\model$ in terms of different hyperparameters (Section~\ref{hyperparameter sensitivity analysis})?
    \item \textbf{RQ4:} What are the final recommendation results of $\model$ from a qualitative perspective (Section~\ref{Case study})?
\end{itemize}

\subsubsection{Comparison of overall performance (RQ1)}
\label{Comparison of overall performance}
Table~\ref{result-three-datasets} presents a comprehensive performance comparison between $\model$ and state-of-the-art GNN-based methods using the evaluation metrics $Precision@K$, $Recall@K$, and $nDCG@K$ with varying values of $K$. Across all four datasets (ML-1M, Amazon-Book, Yelp, and KuaiRec), $\model$ consistently outperforms the five baseline methods, demonstrating the success and effectiveness of the designed message-passing approach on both the positive and negative graphs.
Notably, the performance improvement of $\model$ on KuaiRec is particularly significant compared to the other datasets. For instance, $\model$ outperforms the runner-up LightGCN by 0.85\% in terms of $Recall@5$ and 2.87\% in terms of $Recall@10$. This outcome highlights the advantage of $\model$ when dealing with biased datasets where the number of positive ratings is considerably lower than negative ratings.
In comparison to SiReN, which utilizes an attention model to integrate embeddings from the positive and negative graphs, $\model$ surpasses it in empirical evaluation. It is because $\model$ generates the $\dislike$ embedding $\mathbf{V}$ from the negative graph, which provides a comprehensive user profile and enables the filtering of irrelevant items.
Interestingly, SGCN, which relies on the balance theory assumption, performs poorly compared to other methods. This finding suggests that the balance theory assumption, designed for signed unipartite graphs, is not suitable for real-world recommendation scenarios where users typically have diverse interests.

\subsubsection{Ablation studies (RQ2)}
\label{Ablation studies}
The ablation studies on $\model$ are conducted to investigate the functions of different components. Four variants of $\model$ are designed and evaluated:

\begin{itemize}[leftmargin=*]
\item \textbf{Variant-A}: Using message passing on the negative graph $\mathcal{G}_{n}$.
\item \textbf{Variant-B}: Using message passing on the positive graph $\mathcal{G}_{p}$.
\item \textbf{Variant-C}: Using message passing on both $\mathcal{G}_{p}$ and $\mathcal{G}_{n}$.
\item \textbf{Variant-D}: Introducing graph contrastive learning on Variant-C.
\end{itemize}

The results of the ablation studies on the ML-1M and KuaiRec datasets are presented in Table~\ref{ablation-study}. The observations from the ablation studies are as follows.

\textbf{Variant-A}: Variant-A, which only uses message passing on the negative graph $\mathcal{G}_{n}$, exhibits poor performance in all metrics on both datasets. It indicates that positive feedback is crucial for recognizing users' interests, and negative feedback alone cannot replace it, although it helps recognize users' dislikes.

\textbf{Variant-B} vs. \textbf{Variant-C}: Comparing Variant-B (message passing only on $\mathcal{G}{p}$) and Variant-C (message passing on both $\mathcal{G}{p}$ and $\mathcal{G}_{n}$), it is observed that Variant-C, which integrates the structural information from the negative graph, performs better. It suggests that incorporating the negative graph enhances the model's performance.

\textbf{Variant-C} vs. \textbf{Variant-D}: Introducing the contrastive learning loss on $\mathcal{G}_{n}$ in Variant-D further improves the model's performance.
For instance, Variant-D achieves a 3.14\% higher $Recall@10$ than Variant-C on the KuaiRec dataset.
It demonstrates the effectiveness of contrastive learning for learning accurate $\dislike$ embeddings from the negative graph.

\textbf{Variant-D} vs. \textbf{$\model$}: Comparing Variant-D and the full $\model$, it is observed that leveraging the disinterest-score filter in ranking consistently improves the performance of Variant-D. It confirms the accuracy of $\dislike$ scores and the effectiveness of the disinterest-score filter.

% \begin{figure}[t]
% \centering
% \setlength{\abovecaptionskip}{-0.0cm}
% \setlength{\belowcaptionskip}{-0.2cm}
% \hspace{0.0in}
% \subfigure[\scriptsize{\textcolor{black}{1.}}]{
% \includegraphics[width=0.14\textwidth]{figures/ghcf_precision.png}
% \hspace{-0.0in}}
% \subfigure[\scriptsize{\textcolor{black}{2.}}]{
% \includegraphics[width=0.14\textwidth]{figures/ghcf_recall.png}
% \hspace{-0.0in}}
% \subfigure[\scriptsize{\textcolor{black}{3.}}]{
% \includegraphics[width=0.14\textwidth]{figures/ghcf_ndcg.png}
% \hspace{0.0in}}
% \caption{\textcolor{black}{Comparison of methods for modeling negative feedback using different strategies on ML-1M.}}
% \label{multi_behavior_fig}
% \vspace*{-3mm}
% \end{figure}

\begin{figure*}[t]
\centering
\setlength{\abovecaptionskip}{-0.0cm}
\setlength{\belowcaptionskip}{0.0cm}
\subfigure[\scriptsize{\textcolor{black}{Embedding size.}}]{
\includegraphics[width=0.20\textwidth]{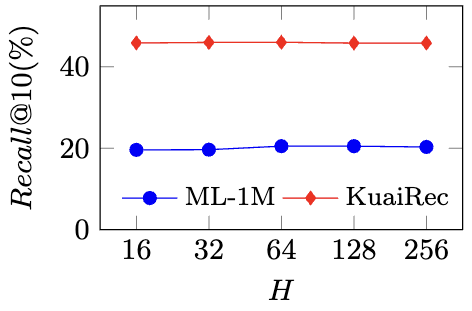}
\hspace{0.1in}}
\subfigure[\scriptsize{\textcolor{black}{Layer number of GNNs.}}]{
\includegraphics[width=0.20\textwidth]{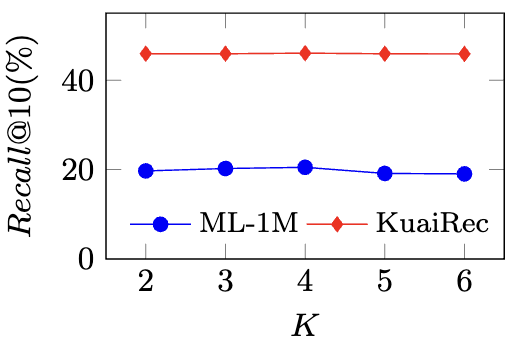}
\hspace{0.1in}}
\subfigure[\scriptsize{\textcolor{black}{Ratio of edge removing.}}]{
\includegraphics[width=0.20\textwidth]{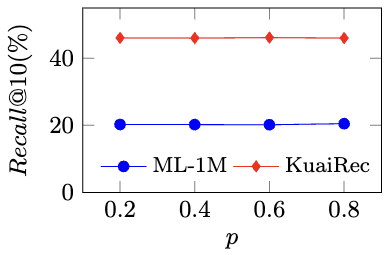}
\hspace{0.1in}}
\subfigure[\scriptsize{\textcolor{black}{Feedback-aware coefficient.}}]{
\includegraphics[width=0.20\textwidth]{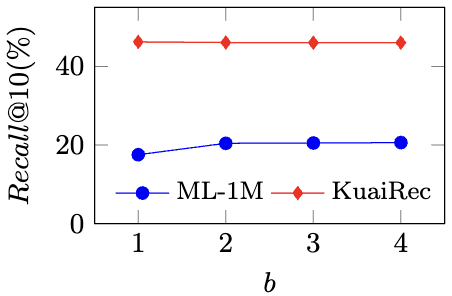}
\hspace{0.0in}}
\subfigure[\scriptsize{\textcolor{black}{Filter score.}}]{
\includegraphics[width=0.20\textwidth]{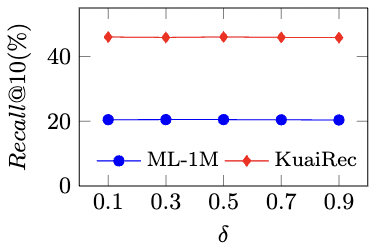}
\hspace{0.1in}}
\subfigure[\scriptsize{\textcolor{black}{Contrastive learning coefficient.}}]{
\includegraphics[width=0.20\textwidth]{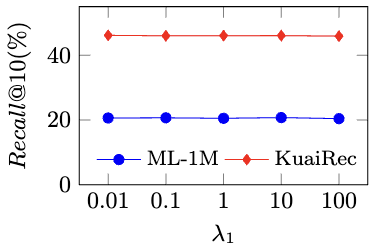}
\hspace{0.1in}}
\subfigure[\scriptsize{\textcolor{black}{Regularization coefficient.}}]{
\includegraphics[width=0.20\textwidth]{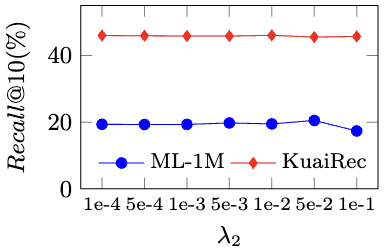}
\hspace{0.1in}}
\subfigure[\scriptsize{\textcolor{black}{Temperature coefficient.}}]{
\includegraphics[width=0.20\textwidth]{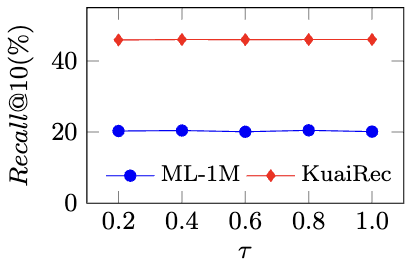}
\hspace{0.0in}}
\caption{\textcolor{black}{Results of sensitivity analysis on ML-1M and KuaiRec.}}
\label{result_sensitivity_analysis}
\vspace*{-3mm}
\end{figure*}

\begin{figure}[t]
\centering
\setlength{\abovecaptionskip}{-0mm}
\setlength{\belowcaptionskip}{-4mm}
\hspace{0.0in}
\subfigure[\scriptsize{\textcolor{black}{Liked videos in training set.}}]{
\includegraphics[width=0.19\textwidth]{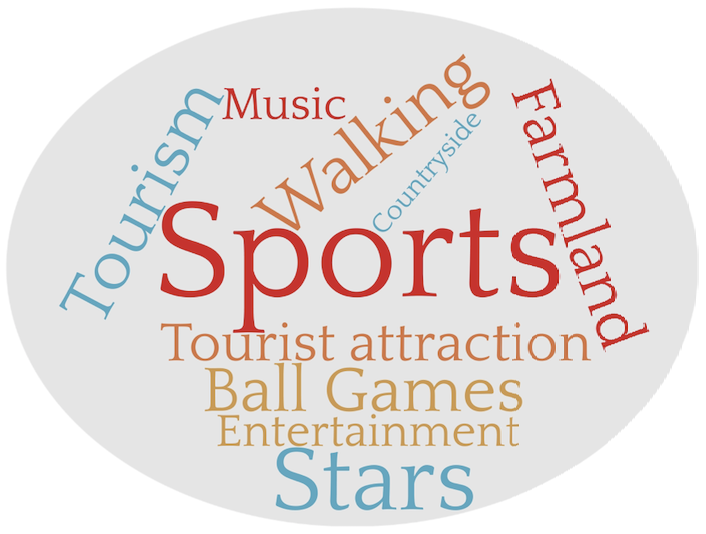}
\hspace{0.2in}}
\subfigure[\scriptsize{\textcolor{black}{Disliked videos in training set.}}]{
\includegraphics[width=0.19\textwidth]{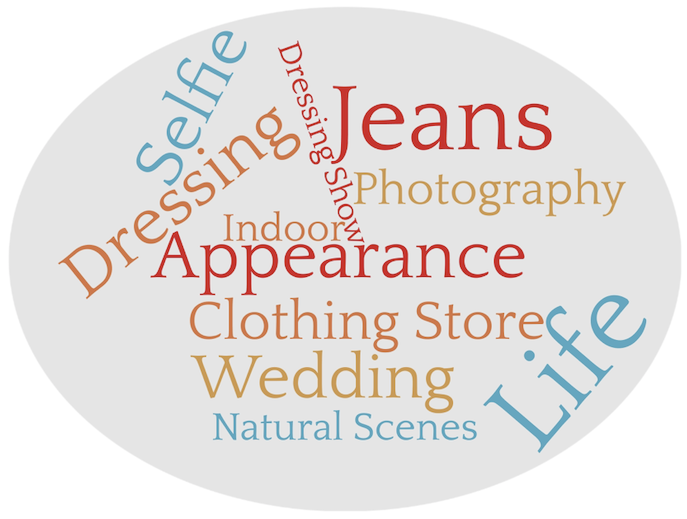}
\hspace{0.0in}}
\subfigure[\scriptsize{\textcolor{black}{Recommended videos before filtering.}}]{
\includegraphics[width=0.19\textwidth]{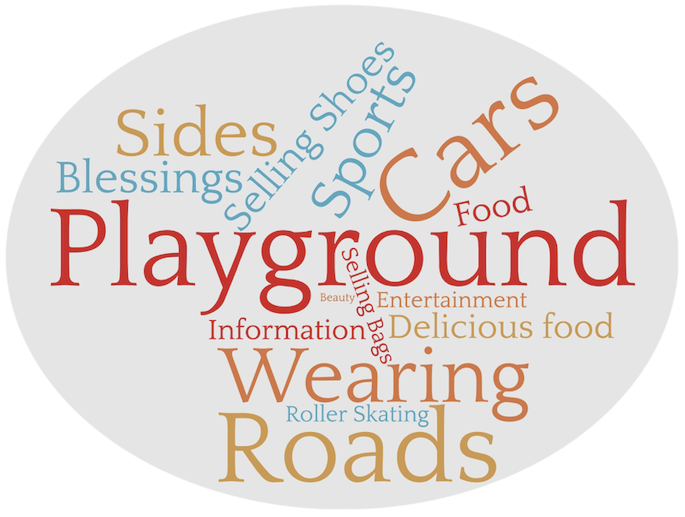}
\hspace{0.2in}}
\subfigure[\scriptsize{\textcolor{black}{Recommended videos after filtering.}}]{
\includegraphics[width=0.19\textwidth]{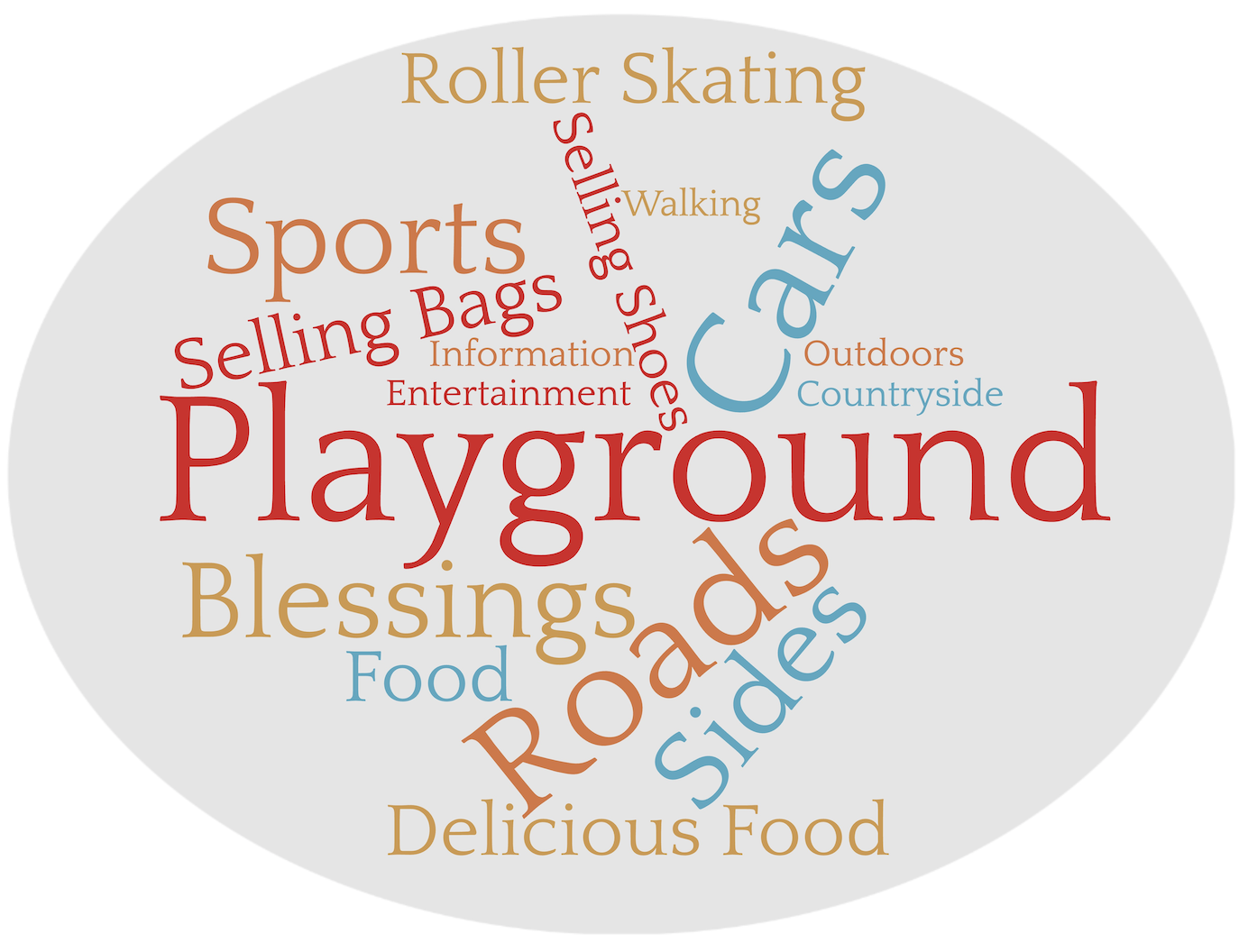}
\hspace{0.0in}}
\caption{\textcolor{black}{Tag clouds of a specific user on the KuaiRec dataset. Each figure presents the tags of the top-10 videos.}}
\label{case_study_fig}
\vspace*{-1mm}
\end{figure}

% \vspace*{-3mm}
% \subsubsection{Comparison with the multi-behavior model (RQ3)}
% \label{Comparison with multi-behavior model}
% We conducted a comparison of three strategies for utilizing negative feedback in recommendation models. These strategies include: not using negative feedback (e.g., NGCF~\cite{NGCF}), constructing a multi-relational graph based on both positive and negative feedback (e.g., GHCF~\cite{GHCF}), and our proposed message passing on both positive and negative feedback. The comparison results are presented in Figure~\ref{multi_behavior_fig}.
% We can see that GHCF performs worst in this comparison, which indicates that directly using negative feedback as another type of relation in the graph cannot unleash its advantages and even has a negative impact on NGCF. Using message passing on both positive and negative feedback 

\subsubsection{Hyperparameter sensitivity analysis (RQ3)}
\label{hyperparameter sensitivity analysis}
To evaluate the sensitivity of $\model$ to different hyperparameters, we conduct a comprehensive hyperparameter sensitivity analysis on ML-1M and KuaiRec. We systematically vary the values of key hyperparameters and measure their impact on the model performance in terms of $Recall@10$. The results are shown in Figure~\ref{result_sensitivity_analysis} and the following findings were observed:

\begin{itemize}[leftmargin=*]
    \item GNNs layer number $K$: As shown in Figure~\ref{result_sensitivity_analysis} (b), we observed that the $Recall@10$ metric initially increases with an increasing number of GNNs layers on the ML-1M dataset. However, beyond a certain point, the $Recall@10$ value starts to decrease.
    This observation aligns with the phenomenon of over-smoothing, where an excessive number of GNNs layers can cause the aggregation of node embeddings to become too similar, resulting in the loss of discriminative information. Additionally, as the number of GNNs layers increases, the computational efficiency of the model may be negatively impacted.
    Considering both the risk of over-smoothing and computational efficiency, we recommend setting $K$ as 3 or 4 to ensure good recommendation outcomes while maintaining computational efficiency.
    \item Feedback-aware coefficient $b$: From the analysis of Figure~\ref{result_sensitivity_analysis} (d), we observed that $b=1$ resulted in inferior performance compared to other values of $b$ on ML-1M. It indicates that discriminating between positive and negative feedback during the optimization process is crucial for achieving better results on ML-1M. The suboptimal performance of $b=1$ suggests that the model might not adequately capture the discriminative signals between positive and negative feedback when they are given equal weight.
    On the KuaiRec dataset, the stability of $\model$'s performance and its insensitivity to different values of $b$ suggest that the dataset's inherent characteristics might diminish the significance of distinguishing between positive and negative feedback.
    Based on these observations, we recommend setting $b$ as 2 or 3.
    \item Regularization coefficient $\lambda_{2}$: As shown in Figure~\ref{result_sensitivity_analysis} (g), $\lambda_{2}=0.1$ performs worst compared with others on ML-1M and KuaiRec.
    Although the L2 regularization term in Eq.~(\ref{total_loss}) can prevent overfitting, high $\lambda_{2}$ excessively penalizes the model's parameters, resulting in underfitting. Hence, we suggest selecting $\lambda_{2}$ from the range of [0.01, 0.05] for $\model$.
    \item Others: We found that $\model$ demonstrates robustness to various other hyperparameters, including the edge removing ratio $p$ and contrastive learning coefficient $\lambda_{1}$.
\end{itemize}

\subsubsection{Case study (RQ4)}
\label{Case study}
In this subsection, we evaluate the recommendation quality of $\model$ by analyzing the tag information of videos in KuaiRec.
In Figure~\ref{case_study_fig} (a), we observe that the user has a preference for outdoor sports-related videos based on the tags of liked videos in the training set. Conversely, Figure~\ref{case_study_fig} (b) displays the tags of disliked videos, indicating disinterest in videos related to dressing or clothing.
Figure~\ref{case_study_fig} (c) and Figure~\ref{case_study_fig} (d) depict the tags of the recommended videos generated by $\model$ before and after the filtering process, respectively.
Our observations reveal the following insights: In Figure~\ref{case_study_fig} (c), the recommended videos generated by $\model$ generally align with the user's interests depicted in Figure~\ref{case_study_fig} (a), except for a few specific words such as ``Wearing'' and ``Beauty''.
With the disinterest-score filter (Figure~\ref{case_study_fig} (d)), $\model$ successfully filters out less relevant recommendations, while suggesting more relevant videos with tags like ``Walking'', ``Outdoors'', and ``Countryside''.
These findings emphasize two key points: 1) $\model$ effectively captures both user interests and disinterests from the training data, and 2) the implementation of the disinterest-score filter proves to be an effective approach for generating more relevant recommendation outcomes.

% \begin{table*}[t]
% \setlength\tabcolsep{3pt}
% \small
% \caption{Case study. Left (right) is the recommended top 10 videos based on positive (negative) feedback.}
% \label{case-study}
% \begin{center}
% \begin{threeparttable}
% \begin{tabular}{cc|cc}
% \toprule[1pt]
% \multicolumn{2}{c|}{\bf Positive feedback} &\multicolumn{2}{c}{\bf Negative feedback}  \\
% \bf Rank &\bf Tag list of video &\bf Rank &\bf Tag list of video \\
% \midrule[1.0pt]
% 1 &['Playground', 'Sports'] &1 &['Dressing Show'] \\
% 2 &['Roads', 'Cars'] &2 &['Self Portrait Face', 'Appearance'] \\
% 3 &['Selling Bags', 'Cars'] &3 &['Side View', 'Style', 'Life'] \\
% 4 &['Selling Shoes', 'Wearing'] &4 &['Self Portrait Face', 'Appearance'] \\
% 5 &['Sides', 'Sports'] &5 &['Front View', 'Dance'] \\
% 6 &['Information'] &6 &['Microphone', 'Entertainment'] \\
% 7 &['Blessings', 'Entertainment'] &7 &['Blur', 'Information'] \\
% 8 &['Food', 'Delicious food'] &8 &['Front View', 'Talent'] \\
% 9 &['Roller Skating', 'Sports'] &9 &['Chicken Eating Game', 'Fun'] \\
% 10 &['Selling Shoes', 'Beauty'] &10 &['Women Pure Outdoor', 'Appearance'] \\
% \bottomrule[1pt]
% \end{tabular}
% \end{threeparttable}
% \end{center}
% \end{table*}

\section{Conclusion and Future Work}
\label{Conclusion and Future Work}
% In this work, we address the problem of leveraging negative feedback to improve recommender systems. Existing approaches in the literature focused on GNN-based recommendation models that only consider message passing on the positive graph. To overcome this limitation and capture high-order structural information from both positive and negative graphs, we propose a novel GNN-based recommendation model called $\model$.
% By aggregating and updating messages on these two graphs, we enable the model to effectively incorporate positive and negative feedback.
% Experimental evaluations conducted on four real-world datasets demonstrate that $\model$ consistently outperforms state-of-the-art GNN-based recommendation methods.
% In future work, we plan to investigate the exposure bias issue in GNN-based recommendation models.

In this work, we address the problem of leveraging negative feedback to improve recommender systems. Existing approaches in the literature focused on GNN-based recommendation models that only consider message passing on the positive graph. To overcome this limitation and capture high-order structural information from both positive and negative graphs, we propose a novel GNN-based recommendation model called $\model$.
By aggregating and updating messages on these two graphs, we enable the model to effectively incorporate positive and negative feedback. Additionally, we employ contrastive learning on the negative graph to reduce noise and filter out items with high $\dislike$ scores, ensuring the relevance of the recommended results.
Experimental evaluations conducted on four real-world datasets demonstrate that $\model$ consistently outperforms state-of-the-art GNN-based recommendation methods. We also conduct an in-depth analysis of $\model$ to validate its effectiveness across different components and its robustness to hyperparameters.
In the future, we plan to investigate the exposure bias issue in GNN-based recommendation models.

\newpage
\bibliographystyle{ACM-Reference-Format}
\bibliography{sample-base}

% \appendix

% \section{XXX}

\end{document}